\documentclass[fleqn,usenatbib,mn2e,article]{mnras}
\usepackage{newtxtext,newtxmath,subcaption}
\usepackage[T1]{fontenc}
\usepackage{ae,aecompl}
\usepackage{amsfonts, natbib, txfonts,epsfig,rotating,txfonts,mathtools,supertabular,booktabs,multirow,geometry,longtable,mathptmx,xfrac, txfonts,verbatim, latexsym}
\usepackage[utf8]{inputenc}
\usepackage{pdflscape}
\usepackage{graphicx,tabularx}

\usepackage{graphicx}	
\usepackage{amsmath}	
\usepackage{amssymb,units}	
\usepackage{pdflscape,tabularx}
\usepackage{gensymb, esint}
\def\Plus{\texttt{+}}
\def\Equ{\texttt{=}}
\def\Minus{\texttt{-}}
\newcommand{\angstrom}{\mbox{\normalfont\AA}}

\newcommand{\dd}{\mathop{}\!\mathrm{d}}

\title{AGN orientation through the spectroscopic correlations and model of dusty cone shell}

\author[M. Laki\'{c}evi\'{c} et al.]{
M. Laki\'{c}evi\'{c}$^{1}$\thanks{E-mail: mlakicevic@aob.rs},  
J. Kova\v{c}evi\'{c}-Doj\v{c}inovi\'{c}$^{1,2}$ and
L. \v{C}. Popovi\'{c}$^{1,3}$
\\
$^{1}$Astronomical Observatory Belgrade; Volgina 7, 11060 Belgrade, Serbia\\
$^{2}$Isaac Newton Institute of Chile, Yugoslavia Branch\\
$^{3}$Department of Astronomy, Faculty of Mathematics, Belgrade University\\
}

\date{Received YYY}

\pubyear{2021}

\begin{document}
\label{firstpage}
\pagerange{\pageref{firstpage}--\pageref{lastpage}}
\maketitle

\begin{abstract}
The differences between Narrow Line Seyfert 1 galaxies (NLS1s) and Broad Line AGNs (BLAGNs) are not completely understood; it is thought that they may have different inclinations and/or physical characteristics. The FWHM(H$\beta$)--luminosities correlations are found for NLS1s and their origin is the matter of debate. Here we investigated the spectroscopic parameters and their correlations considering a dusty,   cone model of AGN. We apply a simple conical dust distribution (spreading out of broad line region, BLR), assuming that the observed surface of the model is in a good correlation with MIR emission. The dusty cone model in combination with a BLR provides the possibility to estimate luminosity dependence on the cone inclination. The FWHM(H$\beta$)--luminosities correlations obtained from model in comparison with observational data show similarities which may indicate the influence of AGN inclination and structure to  this correlation. An alternative explanation for FWHM(H$\beta$)--luminosities correlations is the selection effect by the black hole mass. These FWHM(H$\beta$)--luminosities correlations may be related to the starburst in AGNs, as well. The distinction between spectral properties of the NLS1s and BLAGNs could be caused by multiple effects: beside physical differencies between NLS1s and BLAGNs (NLS1s have lighter black hole mass than BLAGNs), inclination of the conical AGN geometry may have important role as well, where NLS1s may be seen in lower inclination angles.
\end{abstract}




\section{Introduction}\label{sec:intro}

A broad line region (BLR) in active galactic nuclei (AGNs) is believed to be formed close to the black hole \citep[several hundreds of light days in radius;][]{Sturm18} where the gas velocity is up to several thousands km s$^{-1}$. The BLR emits broad lines which are mainly broadened due to the violent emitting gas motion around a supermassive black hole. The AGN unification model \citep{Antonucci93,Urru95} relies to the existance of the dusty torus of the size of $\sim$10 pc \citep{Tristram11} around the accretion disc. Out of the dusty torus, there is the narrow line region, where the narrow emission lines arise. According to the unification model, the AGNs Type 1 are seen nearly face on (small inclination angles of BLR, $i$) and these objects have broad and narrow lines in their spectra. Unlike them, the AGNs Type 2 are seen under larger inclination ($i \approx$ 90$\degree$), their BLR is observed through the torus, therefore their BLR is not seen, only narrow emission lines can be observed in their spectra and they are known as obscured AGNs. Some recent mid-infrared (MIR) studies require the polar cones of dust, stretching out from the central source \citep{Honig19} instead (or together with) preveously presumed dusty torus. Recently, MIR observations resolved dusty and molecular structures in polar directions, on scales from 10 to 100 pc \citep{Garcia14,Gallimore16,Alonso-Herrero18,Asmus2019,Alonso-Herrero21,Buat21,Toba21}, while the theoretical works confirmed that these structures may be from dusty wind driven by the radiation pressure \citep[see][and references therein, hereafter S19]{Stalevski19}. The observed AGN outer half-opening cone angles, $\theta \Equ$30-60$\degree$ \citep{Bae16} are similar to torus presumed half-opening angles of 30-60$\degree$ \citep{Marin15}. Finding that the MIR emission in AGNs is probably from dust embedded in polar outf{l}ows in the narrow line region instead of in the torus \citep{Zhang13} may change much in the AGN understanding, as well as in the comprehension of the differences between Narrow line Seyfert 1 (NLS1) galaxies and Broad Line AGNs (BLAGNs).

\citet{Osterbrock85} and \citet{Goodrich89} defined a new class of AGN, NLS1, that have full width at half maximum of broad permitted line H$\beta$, FWHM(H$\beta$) $<$2000 km s$^{-1}$ and flux ratio of total [O\,III]$\lambda$5007 to total H$\beta <$ 3. These AGNs have notable distinctions from BLAGNs, such as narrower broad emission lines, lighter black hole mass (M$_{\rm BH}$) and lower luminosities \citep{Sani10,Schmidt16}, higher polycyclic aromatic hydrocarbon (PAH) presence, denser BLR clouds, possibly lower inclination, more star formation, have more bars and dust spirals, higher accretion rate, strong optical and UV Fe\,II emission, weak [O\,III], lower optical variability and more rapid X-ray variability \citep{Boroson92,Grupe99,Rakshit17b,Laki18}. Some studies suggest that NLS1s and BLAGNs have different geometry \citep{Baldi16,Liu16}. On the other hand, there are indications that NLS1s are regular Seyfert galaxies at an early stage of evolution where black hole is still growing \citep{Mathur2000,Mathur2001,Jin12a}, and that NLS1s possibly have the same characteristic as BLAGNs, but they are only seen under the different inclination angles \citep{Nagao00,Zhang02,Decarli08,Rakshit17a}. Recently there are various proofs that some NLS1s show blazar characteristics \citep{Jarvela15,Yang18}. Nevertheless, \citet{Jarvela17} suggested that NLS1s and BLAGNs have a different enviromental density and distribution, therefore they should not be unified by orientation. Some of foregoing characteristics can indeed be explained by lower inclination and lower M$_{\rm BH}$ of NLS1s, as we discuss in this paper. 

NLS1s show the correlation between FWHM(H$\beta$) and several optical and MIR line and continuum luminosities, while BLAGNs do not have that characteristic \citep[][hereafter La18]{Laki18} and the cause of that is not certain. \citet{Jarvela15} also found somewhat weaker (with a correlation coefficient of $\sim$0.3) trends FWHM(H$\beta$)-luminosities (optical, MIR and radio) for NLS1s. \citet{Popovic11} found that the AGNs with stronger starburst contribution have the correlation between FWHM(H$\beta$) and optical luminosity, while the other AGNs do not have this trend. This finding is similar to the one for NLS1 correlations (see above), as starburst objects are often connected with NLS1s. \citet{Boller01} noticed anticorrelation between soft X-ray excess strength (0.1--2.4  keV) and FWHM(H$\beta$) and between hard continuum slope (2--10 keV) with FWHM(H$\beta$), for NLS1s. \citet{Jarvis06} noted strong correlation between radio spectral index (known to be connected with the source orientation) and FWHM of broad lines H$\beta$ and Mg\,II. AGNs with FWHM(H$\beta$)$>$4000 km s$^{-1}$ have a larger redshifted very broad line region component of H$\beta$ (velocities$\sim$5000 km $^{-1}$) \citep[see][]{Sulentic02,Kovacevic10}. 

If there is not any dust extinction (absorption and/or scaterring) within an AGN, then the angle of view of AGN would not be important since all photons would arrive to the observer. The angle-dependent obscuration is needed for AGN unification understanding \citep{Honig19}. In estimation of the luminosity received from the certain object under different $i$, we are interested in its geometry, size of the observed surface and optical depth. Here we tested if the inclination, observed surface and the optical depth could determine the luminosity of AGNs. 

Inclination of AGNs may not only influence the type of AGNs we detect, but also significantly affects the estimation of M$_{\rm BH}$ when FWHM of broad emission lines is used, as well as it affects some reverberation-based M$_{\rm BH}$ calculations, especially for the objects with the narrowest emission lines \citep{Collin06}. Even $M-\sigma$* relation (between M$_{\rm BH}$ and stellar velocity dispersion) is contaminated by the inclination influence, as inclination may affect measured velocity dispersions by 30$\%$, and consequently M$_{\rm BH}$ may be affected up to 1 dex, where face-on objects have lower, and edge-on sistematicaly higher $\sigma$*, due to contamination by disc stars \citep{Bellovary14}.

The aim of this paper is to explain spectral characteristics and correlations between spectral parameters of the AGNs by the inclination calculated from spectroscopic parameters in combination with the specific conical distribution of dust which is observed in some AGNs. For that purpose, we involve dusty cone models and we test the changes in spectral properties for different angles of view. 

In Section~\ref{sec:Data} we explain the dataset and the used procedure. In Section~\ref{sec:Results}, the major plots and results are shown. In Section~\ref{sec:Disc} the implications of obtained relations are discussed. In Section~\ref{sec:Conclusion}, the concluding remarks are listed. Cosmological parameters used in this work are $\Omega_{m} \Equ$0.3, $\Lambda \Equ$ 0.7 and H$_{0} \Equ$ 70 km s$^{-1}$ Mpc$^{-1}$.

\section{Data and Methods} \label{sec:Data}

\begin{figure}  
\rotatebox{0}{
\includegraphics[width=88mm]{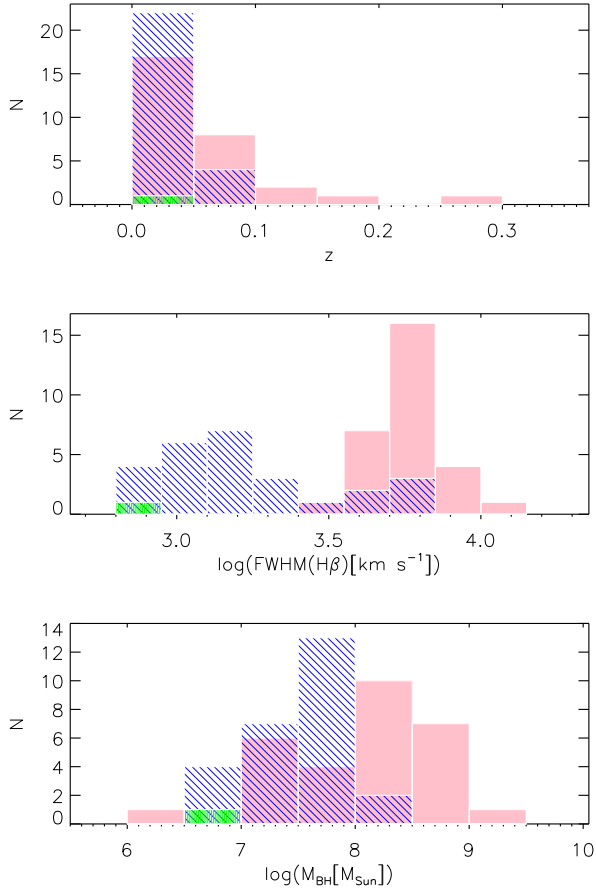}}
\caption{The histograms of major spectral parameters for the AGN dataset used in this paper. The histograms for objects from \citet{Afanasiev19} are coloured with pink, the histograms for objects from \citet{Zhang02} are filled with blue lines, while the object from \citet{Shapovalova12} is filled with green. \label{fig:his}}
\end{figure}

For this research we needed the sample of the AGN Type 1 spectra for which: a) M$_{\rm BH}$ is calculated with some method independent of FWHM(H$\beta$) (in order to calculate inclination $i$, see Eq.~\ref{eq:eq1}), b) for which the MIR spectra are available in order to investigate the correlations of the MIR spectral properties with $i$. Therefore, the data used in this investigation is composed from the two datasets of Type 1 AGNs, \citet{Zhang02} and \citet{Afanasiev19} for which there are available certain spectroscopic parameters needed to calculate the inclination. Data from \citet{Zhang02} are chosen with the same purpose, while here that sample is enriched with the data from \citet{Afanasiev19} and one object from \citet{Shapovalova12}. Data from \citet{Afanasiev19} indicated equatorial scattering. M$_{\rm BH}$s are calculated using stellar velocity dispersions or using polarization in broad lines. Additionally, these objects also have available Spitzer spectra. Reduced Spitzer Space Telescope spectra from Infrared Spectrograph (IRS) instrument, in low-resolution ($\sim$60-127) and/or high-resolution ($\sim$600) are downloaded from the Combined Atlas of Sources with Spitzer IRS Spectra, CASSIS\footnote{https://cassis.sirtf.com/. The Combined Atlas of Sources with Spitzer IRS Spectra (CASSIS) is a product of the IRS instrument team, supported by NASA and JPL. CASSIS is supported by the "Programme National de Physique Stellaire" (PNPS) of CNRS/INSU co-funded by CEA and CNES and through the "Programme National Physique et Chimie du Milieu Interstellaire" (PCMI) of CNRS/INSU with INC/INP co-funded by CEA and CNES.} \citep{Lebouteiller11,Lebouteiller15} for whole dataset. The dataset used in this paper consists of 56 (28 NLS1s and 28 BLAGNs) low redshift AGNs for which some spectral characteristics are shown in Fig.~\ref{fig:his}. 

Optical parameters such as orbital velocities ($\sigma_{V}$, estimated from broad lines), FWHM(H$\beta$), and BLR radius (R$_{\rm BLR}$), are taken from aforementioned three papers. There, R$_{\rm BLR}$ is calculated by reverberation mapping, while $\sigma_{V}$ was calculated from FWHM(H$\beta$), such that FWHM(H$\beta$)$\approx$2.355$\sigma_{V}$. 

The M$_{\rm BH}$s of objects from the sample are taken from the aforementioned three papers as well. For the objects from \citet{Afanasiev19} and from \citet{Shapovalova12}, M$_{\rm BH}$s are derived using polarization in broad lines, while for the objects from \citet{Zhang02}, M$_{\rm BH}$s are obtained using stellar velocity dispersion. 

For the objects from \citet{Zhang02} and \citet{Shapovalova12} inclination is calculated in the same way as in \citet{Afanasiev19}, according to the equation:

\begin{equation}  \label{eq:eq1}
 \sin \left(  i \right) \Equ \sigma_{V} \sqrt{\frac{R_{\rm BLR}}{M_{\rm BH} G}},    
\end{equation} where $G$ is gravitational constant. The spectroscopic data used in this paper are given in the Tables \ref{tab1}, \ref{tabsr} and \ref{tab3}. For the data from \citet{Afanasiev19} $\Delta \sigma_{V}$ is taken from that paper, while for the rest of the data it is estimated as 4\%, as that is the most common value in the histogram of the data from \citet{Afanasiev19}. $\Delta$FWHM(H$\beta$) and $\Delta$R$_{{\rm BLR}}$ are taken from \citet{Afanasiev19}, \citet{Zhang02}, or estimated as 4\% and 20\%, respectively (similarly as for $\Delta \sigma_{V}$). $\Delta$log(M$_{\rm BH}$) is from \citet{Afanasiev19} or estimated as 3.5\%, similarly as for $\Delta \sigma_{V}$. For the equability of two datasets, the errors of inclination were found as 

\begin{equation}
\Delta i \Equ \frac{1}{2} tg \left( i \right) \times \left( \frac{\Delta R}{R} \Plus \frac{2 \Delta \sigma_{V}}{\sigma_{V}} \Plus \frac{\Delta M_{\rm BH}}{M_{\rm BH}} \right), 
\end{equation} for the whole sample.  

\begin{table*} 
\caption{The name of object, coordinates and redshifts are given in columns 1, 2, 3 and 4; FWHM(H$\beta$), its reference, $\Delta$FWHM(H$\beta$) and its reference in columns 5, 6, 7 and 8, while equivalent width (EW) of PAH at 7.7 $\mu$m {and its uncertainty is given in columns 9 and 10}. The data are compiled from the datasets from \citet{Afanasiev19} (A19), \citet{Zhang02} (Z02) and \citet{Shapovalova12} (S12). Only a portion of this table is shown here to demonstrate its form and content.  \label{tab1}}
\centering{
\begin{tabular}{|c|c|c|c|c|c|c|c|c|c|}
\hline\hline
 NAME& RA[$\degree$] &Dec[$\degree$] &z &FWHM(H$\beta$)[km s$^{-1}$] &ref &$\Delta$FWHM(H$\beta$)[km s$^{-1}$] & ref &EW7.7[nm] & $\Delta$EW7.7[nm]  \\  
 (1) &(2)&(3)&(4)&(5) &(6)&(7)&(8)&(9) & (10)\\ \hline 
  Mrk\,335        & 1.581339 &  20.202914  &   0.026 &5306 &          A19 & 200.2                              &  A19& 1.0      &  0.2213 \\ \hline
  Mrk\,1501       & 2.629191 &  10.974862  &   0.089 &5617 &          A19 & 195.5                              &  A19& 12.0     &  3.791028 \\ \hline
  Mkn\,1148       & 12.978167&  17.432917  &   0.064 &4724 &          A19 & 353.2                              &  A19& 6.3      &  1.0538955\\ \hline
  I\,Zw1          & 13.395585&  12.69339   &   0.059 &5011 &          A19 & 280.3                              &  A19& 14.0     &  2.32988    \\ \hline \hline                                                                                                            
\end{tabular}}
\\
\smallskip
\end{table*} 

\begin{table*} 
\caption{The name of object, PAH fraction -- RPAH (see the text), its uncertainty, L6, its error, L12, its error, log(M$_{\rm BH}$[M$_{\odot}$]) and its uncertainty and the reference for the mass. Only a portion of this table is shown here to demonstrate its form and content. \label{tabsr}}
\centering{
\begin{tabular}{|c|c|c|c|c|c|c|c|c|c|}
\hline\hline
 NAME &RPAH &  $\Delta$RPAH&  L6[erg s$^{-1}$] &$\Delta$L6[erg s$^{-1}$] & L12[erg s$^{-1}$] &$\Delta$L12[erg s$^{-1}$] &log(M$_{\rm BH}$[M$_{\odot}$])& $\Delta$log(M$_{\rm BH}$[M$_{\odot}$])&ref  \\  
(1) &(2)&(3)&(4)&(5) &(6)&(7)&(8)&(9) & (10)\\ \hline 
  Mrk\,335         &0.002& 0.071  &   8.117$\times$10$^{43}$ & 2.831$\times$10$^{42}$ &7.197$\times$10$^{43}$ &2.516$\times$10$^{42}$& 7.49  & 0.25  &  A19  \\ \hline
  Mrk\,1501        &0.002& 0.138  &   3.783$\times$10$^{44}$ & 2.641$\times$10$^{43}$ &3.502$\times$10$^{44}$ &2.439$\times$10$^{43}$& 8.57  & 0.26  &  A19 \\ \hline
  Mkn\,1148        &6.83$\times$10$^{-4}$& 0.142&   4.236$\times$10$^{43}$ & 2.030$\times$10$^{42}$ &3.396$\times$10$^{43}$ &1.639$\times$10$^{42}$& 8.69  & 0.18  &  A19  \\ \hline
  I\,Zw1           &0.0    & 0.043&   7.214$\times$10$^{44}$ & 1.727$\times$10$^{43}$ &9.264$\times$10$^{44}$ &2.226$\times$10$^{43}$& 7.46  & 0.3   &  A19 \\ \hline
 \hline
\end{tabular}}
\end{table*} 

\begin{table*} 
\caption{The name of object, R$_{{\rm BLR}}$, its reference, $\Delta$R$_{{\rm BLR}}$, its reference, calculated inclination $i$, $\Delta i$, AGN type, reference, luminosity at the peak of the 7.7 $\mu$m feature and its uncertainties are in columns 1-11. The AGN classification is taken from various literature: \citet{M93} (M93), \citet{G99} (G99), \citet{Klimek04} (K04), \citet{N09} (N09), \citet{Bian10} (B10), \citet{Sani10} (S10), \citet{Veron10} (VC10), \citet{Berton15} (B15), \citet{Markowitz14} (Mar14), \citet{M16} (M16). Only a portion of this table is shown here to demonstrate its form and content. \label{tab3}}
\centering{
\begin{tabular}{|c|c|c|c|c|c|c|c|c|c|c|}
\hline\hline
 NAME &R$_{{\rm BLR}}$[l.d.] & ref &$\Delta$R$_{{\rm BLR}}$ [l.d.]& ref &$i$ & $\Delta i$     &Type &ref  & log($\nu$L7.7)[erg s$^{-1}$] & $\Delta$log($\nu$L7.7)[erg s$^{-1}$]   \\  
(1) &(2)&(3)&(4)&(5) &(6)&(7)&(8)&(9) & (10) & (11) \\ \hline 
  Mrk\,335         &15.7   &   A19 & 3.0 &  A19 &   44.5  &0.147      &NLS1 & S10  & 41.661  & 39.045  \\ \hline
  Mrk\,1501        &72.3   &   A19 & 5.4 &  A19 &   28.8  &0.048      &NLS1 & B15  & 41.895  & 41.173  \\ \hline
  Mkn\,1148        &34.3   &   A19 & 0.1 &  A19 &   13.7  &0.021      &BLAGN& VC10 & 40.669  & 39.071  \\ \hline
  I\,Zw1           &18.7   &   A19 & 2.5 &  A19 &   45.4  &0.145      &NLS1 & S10  & --      & --      \\ \hline
 \hline
\end{tabular}}
\end{table*} 

The {\sl deblendIRS}\footnote{http://www.denebola.org/ahc/deblendIRS/} \citep[see][]{Hernan15} script was used for the decomposition of Spitzer spectra to the AGN, PAH and stellar components \citep[for example see Fig. 8 in][]{Laki17}. The MIR spectral parameters, monochromatic luminosities at 6 and 12 $\mu$m, L6 and L12, respectively and RPAH (fractional contribution of PAH component to the integrated 5-15 $\mu$m luminosity) are obtained using {\sl deblendIRS}. The equivalent width (EW) of PAH at 7.7 $\mu$m are measured using Starlink {\sl Dipso} package. The PAHs at 7.7 $\mu$m are chosen among other PAHs from the available spectra because of high intensity and because they do not coincide with significant MIR spectral lines. 

\subsection{The hyperboloid and disc projection surfaces} \label{sec:surfaces}

In order to estimate the luminosity received from an AGN observed from the different inclinations ($i$), we are interested in its observed surface size, its density, opacity and optical depth. The optical depth ($\tau_{\lambda}$) depends on the path through the object ($s$), density ($\rho$) and the absorption coefficient $\kappa_{\lambda}$ (opacity), such that:
\begin{equation}
  \alpha_{\lambda} \Equ \rho \kappa_{\lambda}   \label{eq:2}
\end{equation} and
\begin{equation}
  \tau_{\lambda} \Equ \int \alpha_{\lambda} \dd s \Equ \alpha_{\lambda} s, \label{eq:3}
\end{equation} assuming that the extinction coefficient $\alpha_{\lambda}$ is constant through the path, while $s$ depends on the viewing angle (see below). The observed intensity from the object, $I$, depends on the $\tau_{\lambda}$ as 
\begin{equation}
  I \Equ I_{0} e^{-\tau},   \label{eq:4}
\end{equation} where $I_{0}$ is the initial intensity from that object.

\begin{figure*} 
\centering
\rotatebox{0}{
\includegraphics[width=90mm]{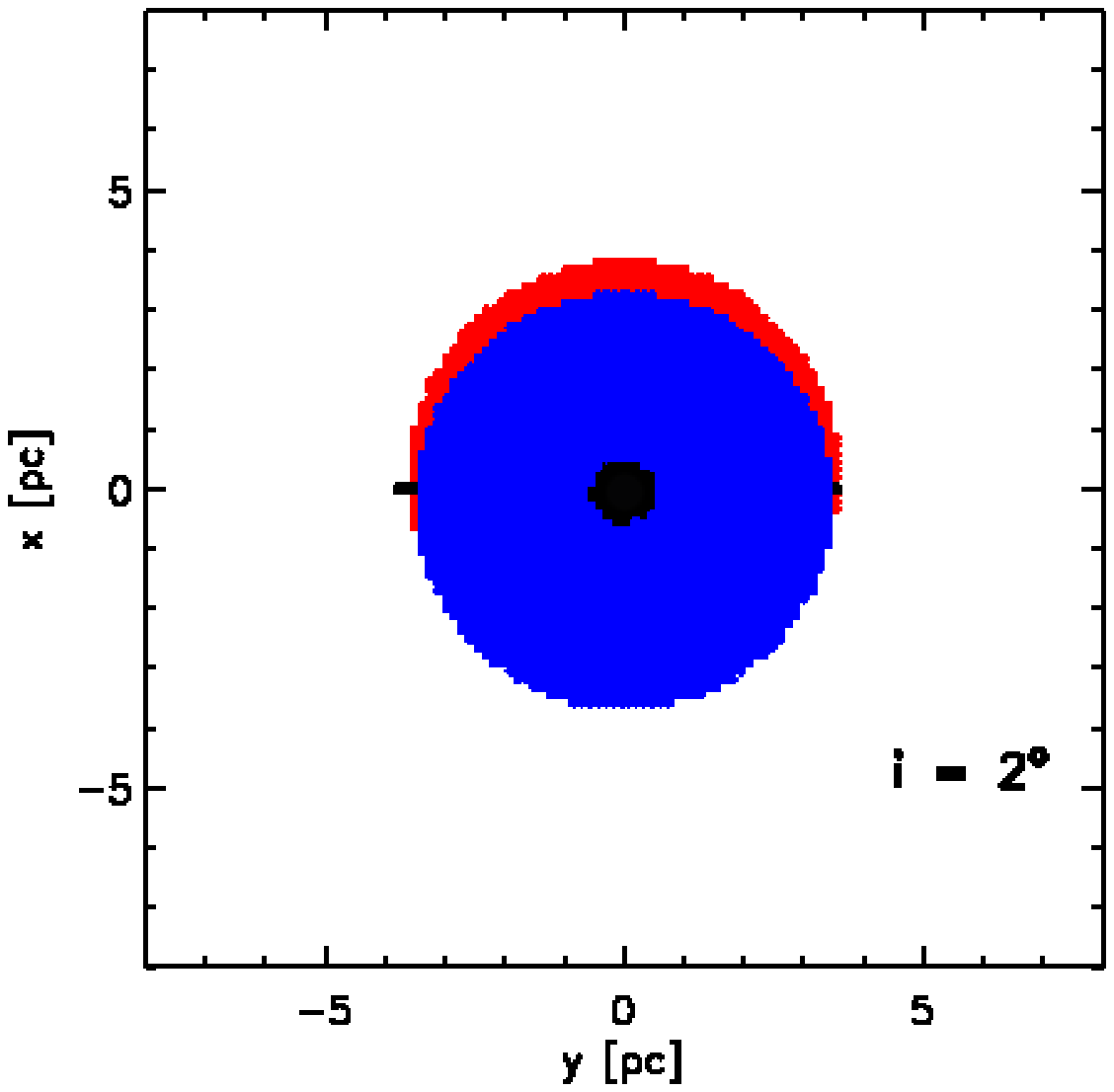}
\includegraphics[width=90mm]{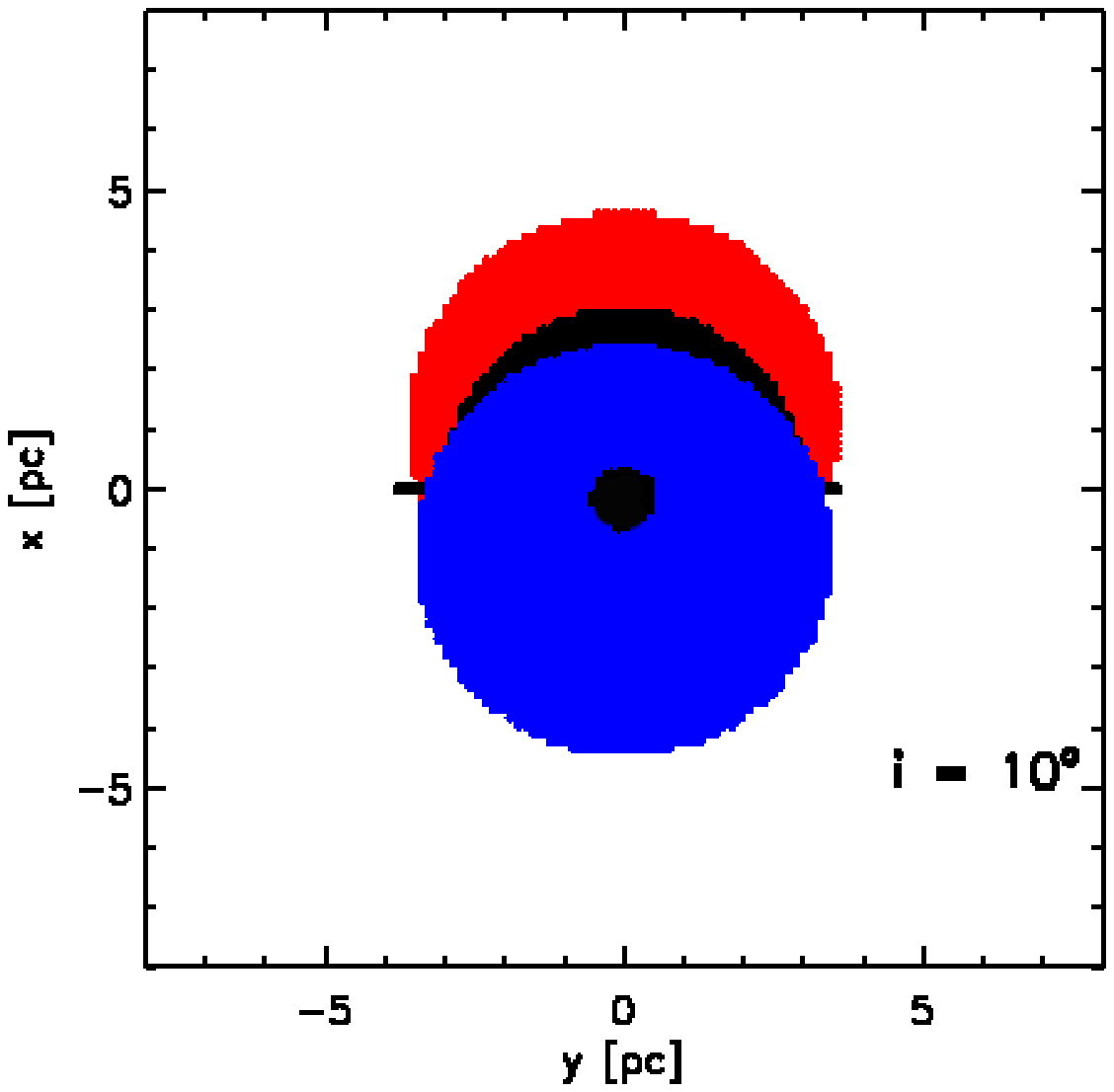}}
\rotatebox{0}{
\includegraphics[width=90mm]{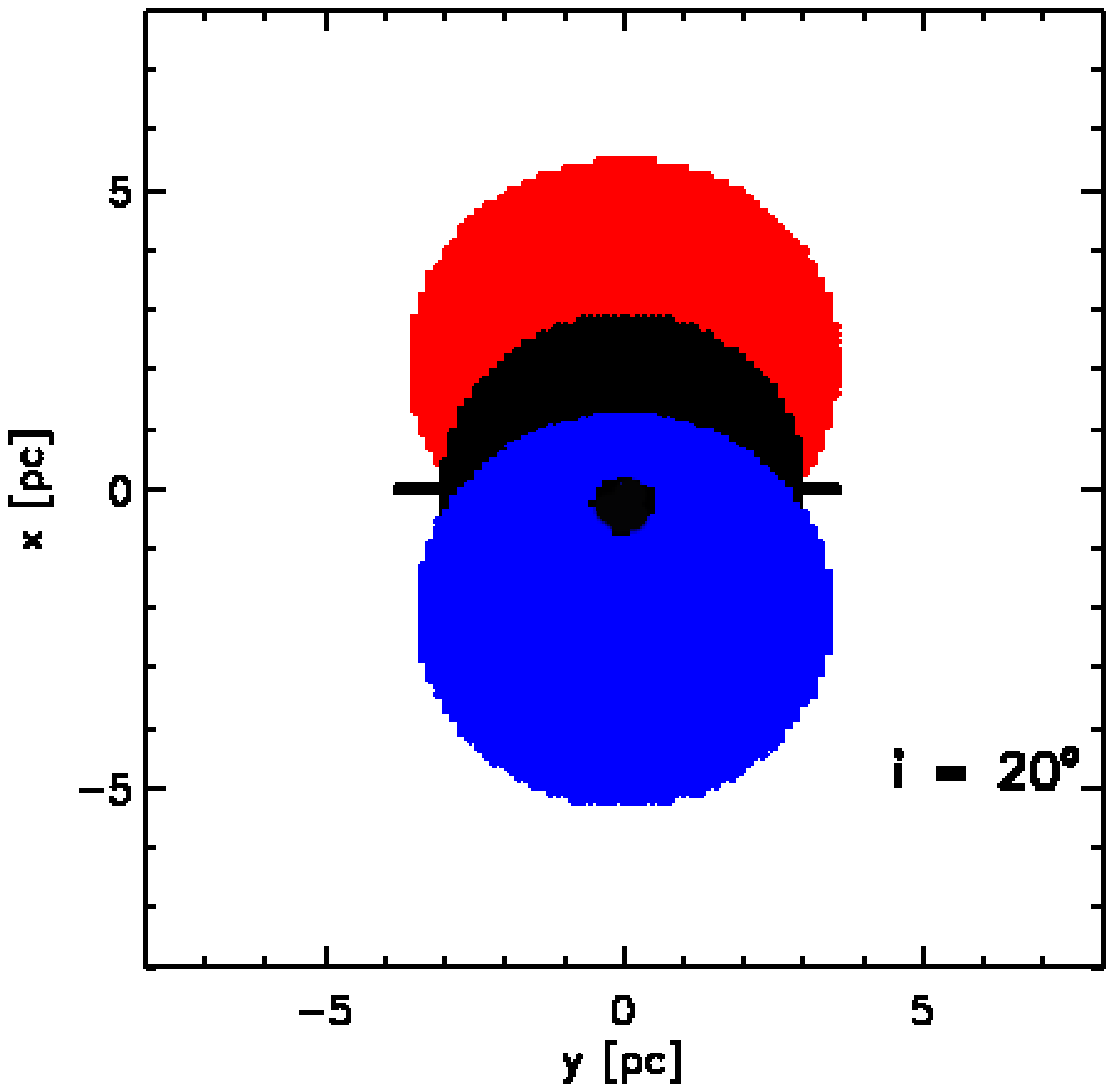}
\includegraphics[width=90mm]{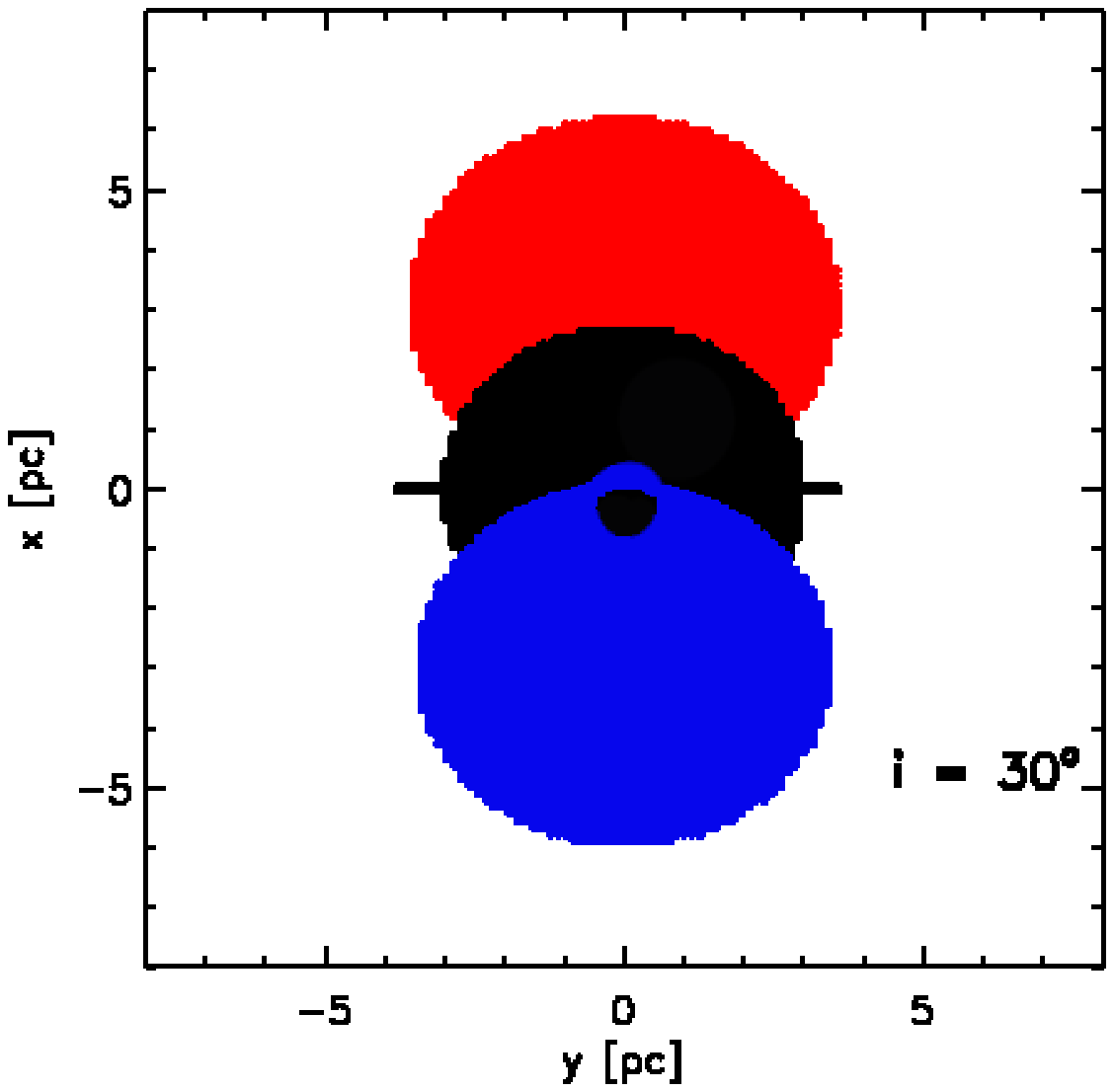}}
\rotatebox{0}{
\includegraphics[width=90mm]{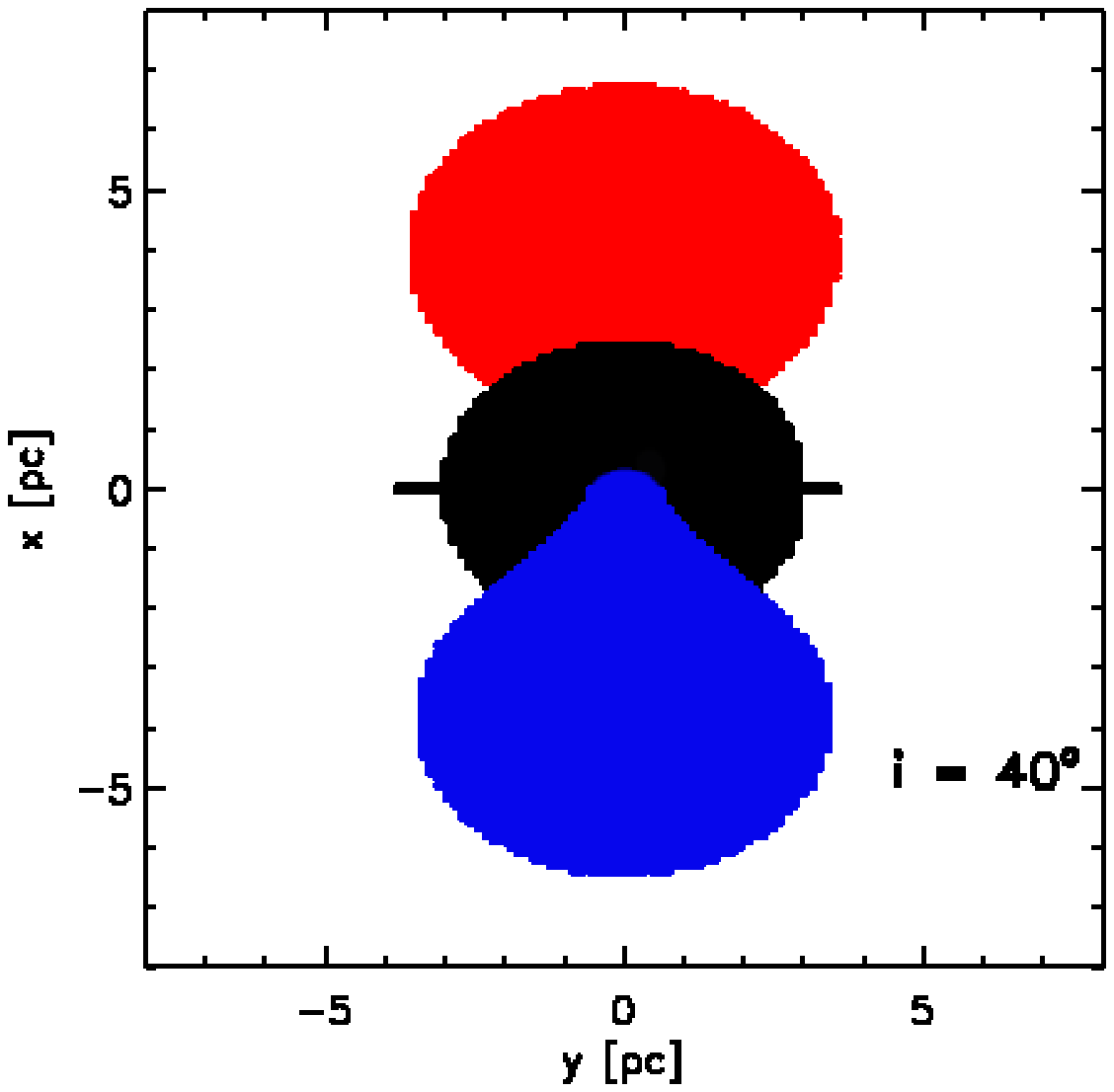}
\includegraphics[width=90mm]{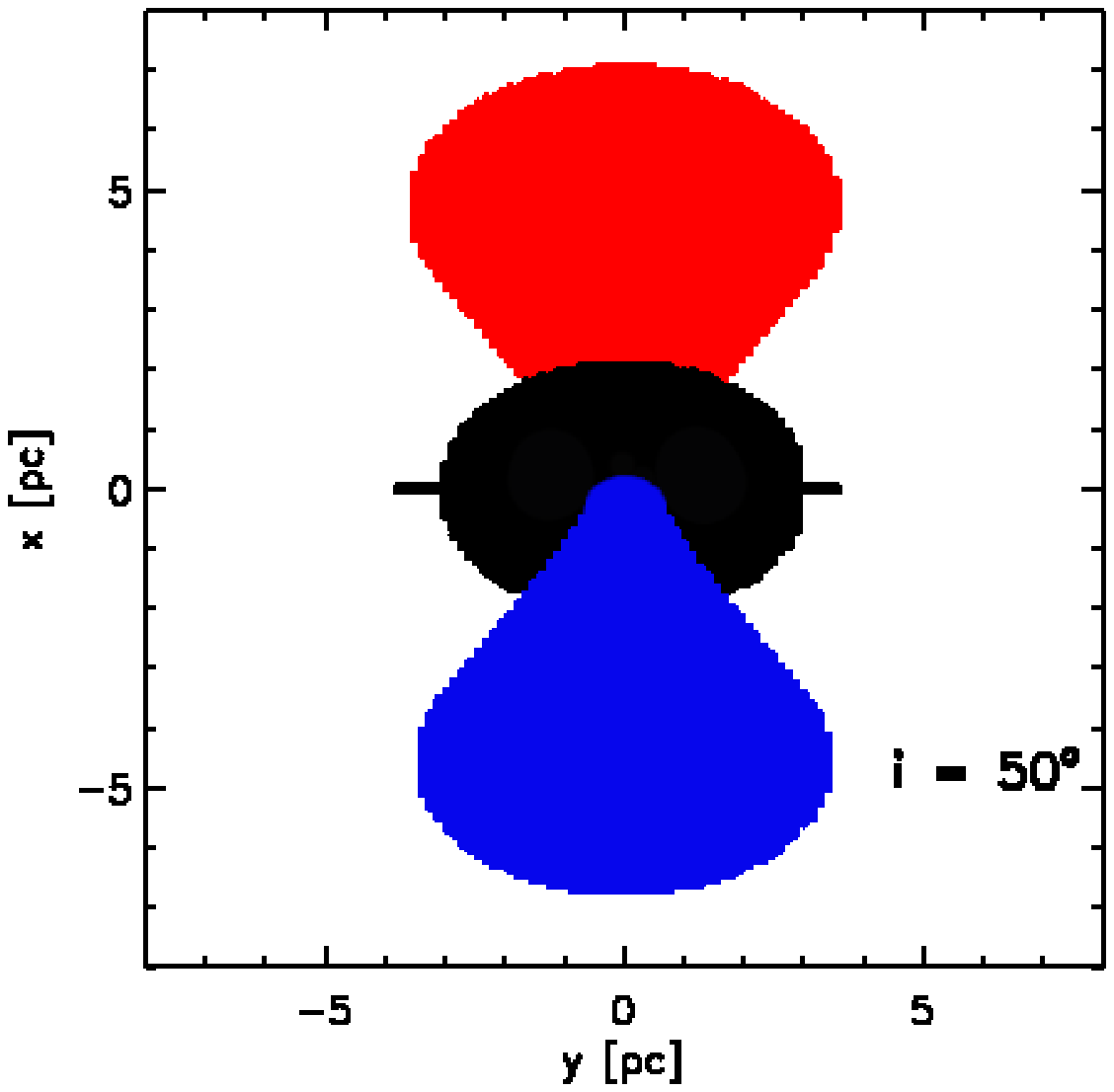}}
\caption{The projections of the polar dusty wind in a form of a hyperboloid shell with thin dusty disc to the plane of observer, for the various inclination angles (written in each panel). The two cones are coloured blue and red, while the dusty disc is black. \label{fig:p}}
\end{figure*}

\begin{figure*} 
\centering
\rotatebox{0}{
\includegraphics[width=90mm]{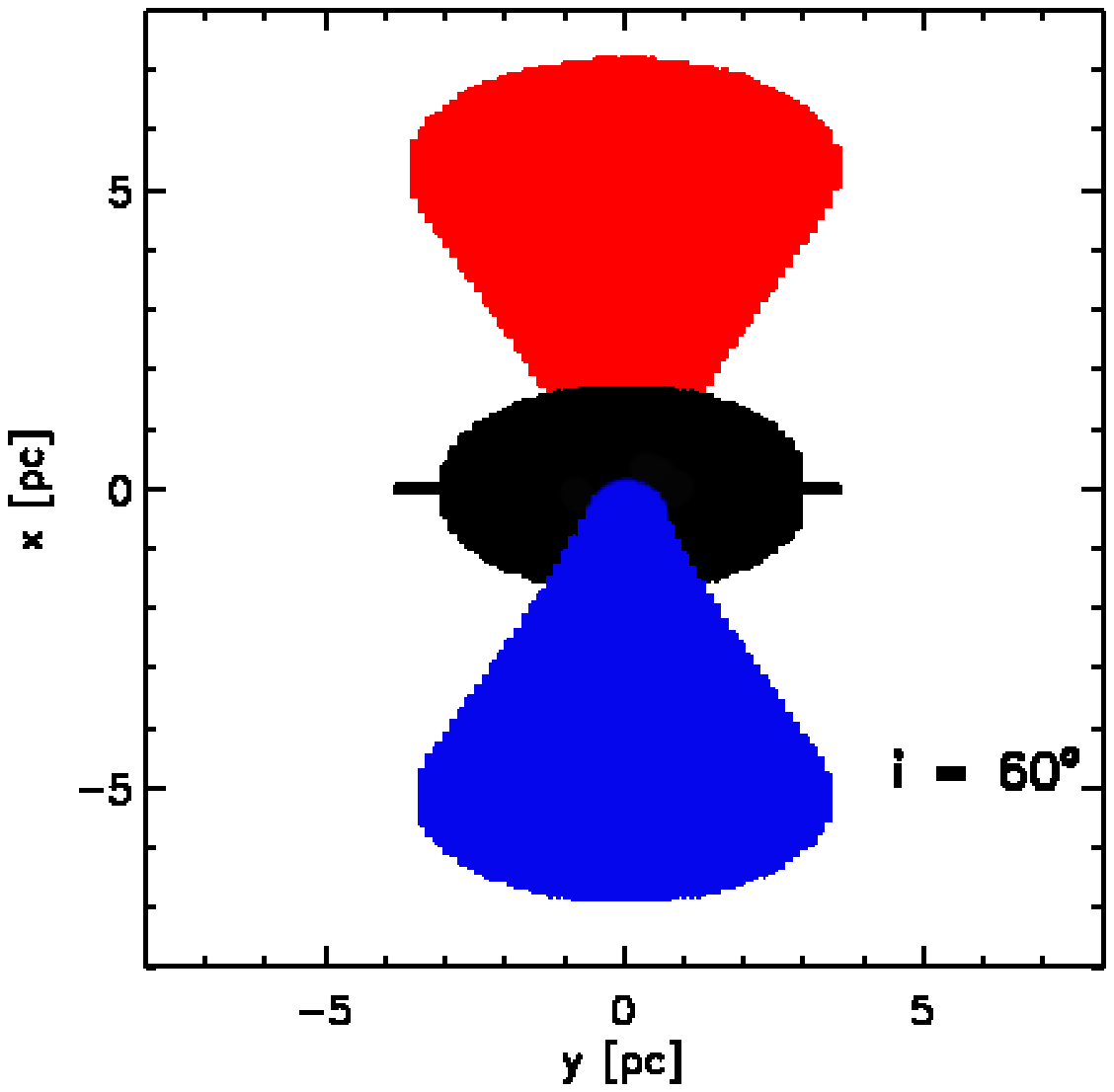}
\includegraphics[width=90mm]{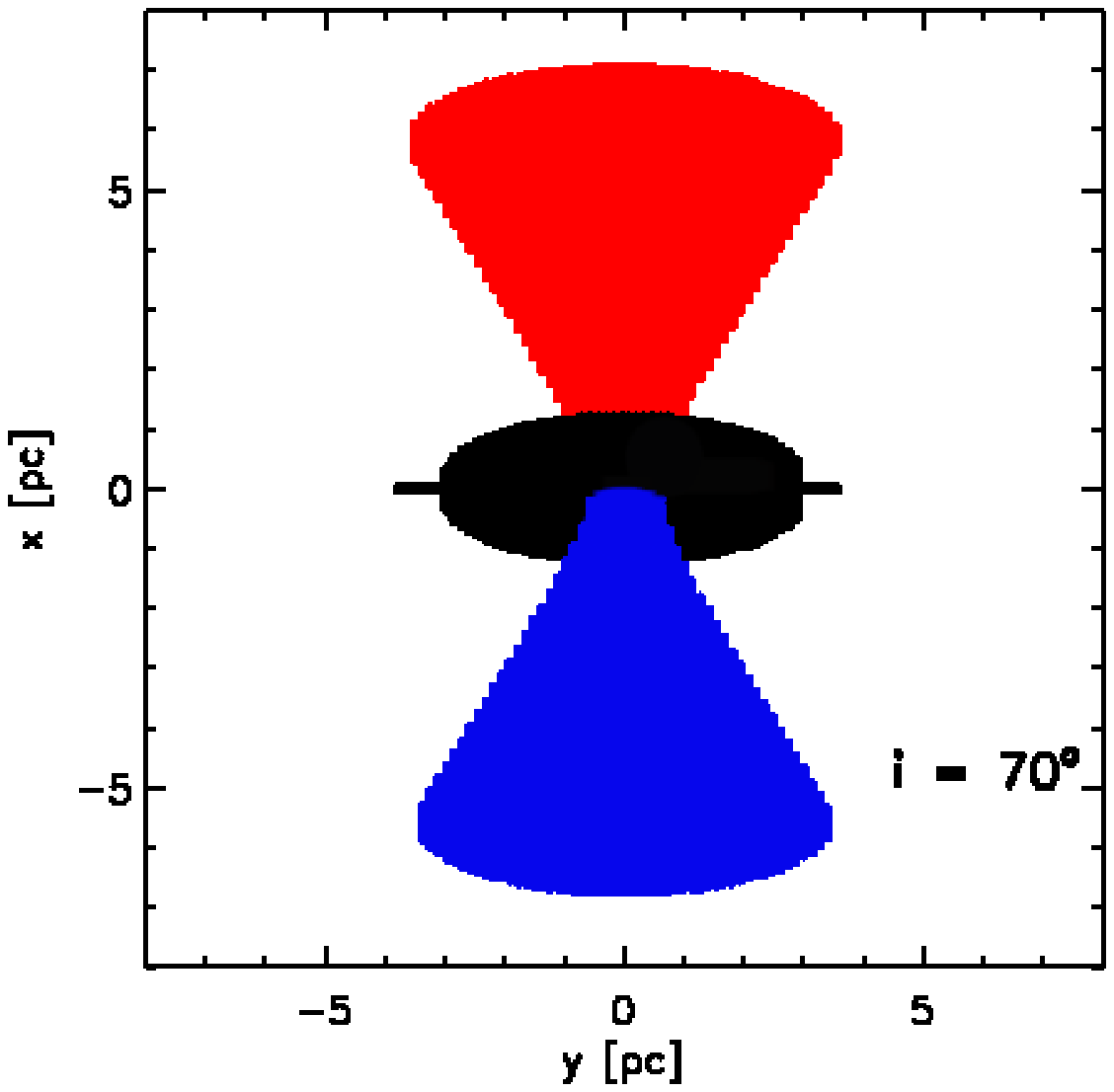}}
\rotatebox{0}{
\includegraphics[width=90mm]{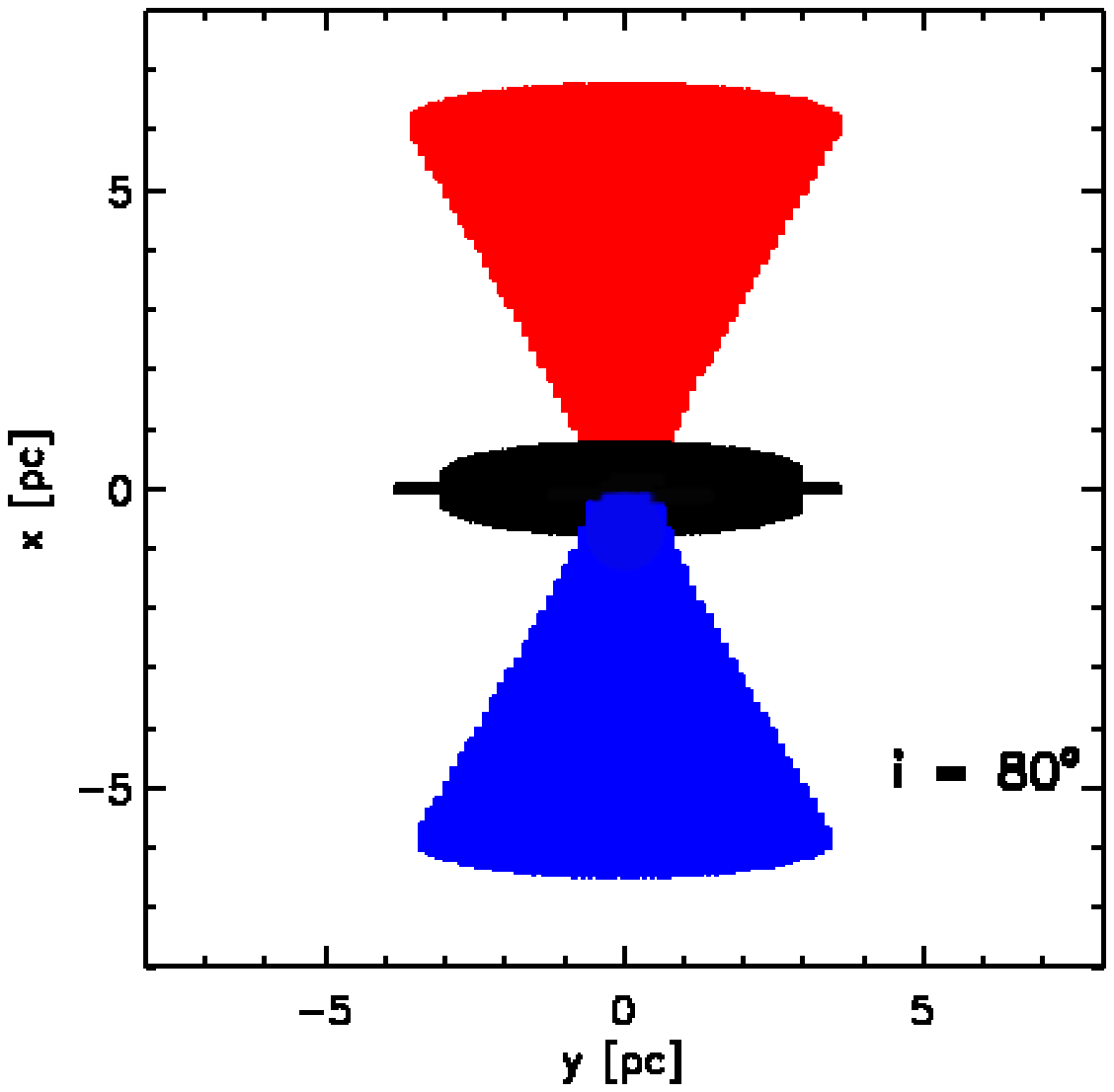}
\includegraphics[width=90mm]{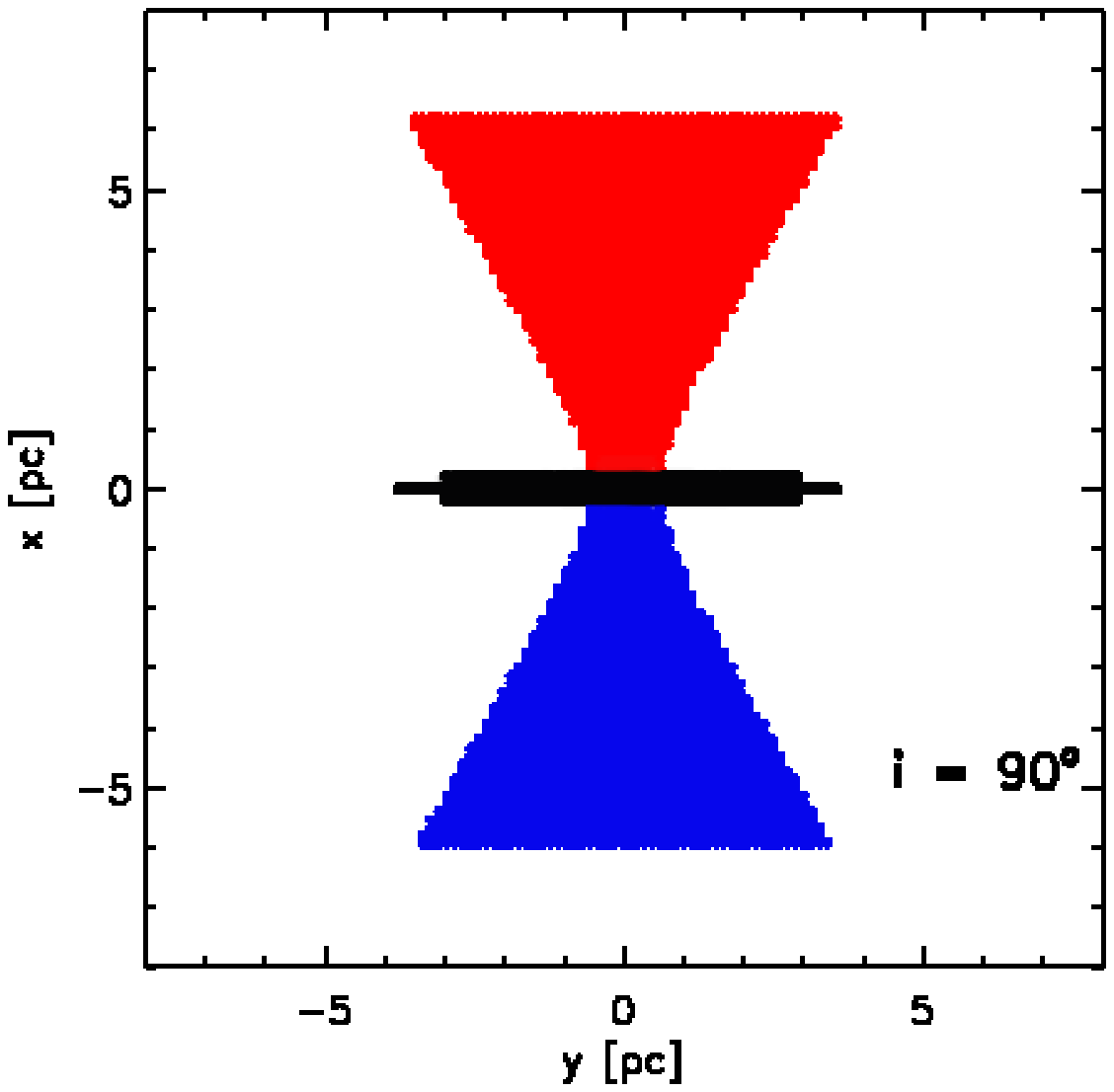}}
\caption{The same as in Fig.~\ref{fig:p}, but on the higher inclinations. \label{fig:p1}}
\end{figure*}

In the case of classical torus model of the AGNs (found based on optical observations) the observed surface of an AGN would decrease with the inclination angle, as the circle of view is becoming the ellipse and luminosities would drop with $i$. Therefore, we used dusty cone models of active galaxies, such as a model made for Circinus\footnote{The Circinus galaxy is one of the closest Type 2 Seyfert galaxies, z$\Equ$0.0014.} galaxy (see S19), as the luminosities have considerable increase with $i$ (for lower inclinations), which may cause FWHM(H$\beta$)--luminosities relations. We check if dusty cone models are related to the correlations between FWHM(H$\beta$) and spectral characteristics. In the model for Circinus galaxy presented in S19 (Section 3.1.1), MIR observations of this AGN are explained as one sheet dusty hyperboloid shell with a thin dusty disc, on parsec scales. This model does not consider dusty torus that is modeled by {\sl SKIRTOR} in S19 (Section 3.1.2), which is one of the most used models in the literature. The conical body is limited with two hyperboloids which have the equations:
\begin{equation}
  \frac{x^{2}}{a^{2}} \Plus \frac{y^{2}}{b^{2}} - \frac{z^{2}}{c^{2}} \Equ 1, 
\end{equation} where $a \Equ b$ (radial symmetry), therefore
\begin{equation}
  \frac{(x^{2} \Plus y^{2})}{a^{2}_{1}} - \frac{z^{2}}{c_{1}^{2}} \Equ 1, 
\end{equation} and
\begin{equation}
  \frac{(x^{2} \Plus y^{2})}{a^{2}_{2}} - \frac{z^{2}}{c_{2}^{2}} \Equ 1, 
\end{equation} where $a_{1} \Equ 0.2$ pc for the inner and $a_{2} \Equ 0.6$ pc for the outer hyperboloid (from S19 model), while parameters $c_{1}$ and $c_{2}$ are 0.35 and 1.04, respectively ($c_{1,2} \Equ a_{1,2}\times$ tan(90$\degree-\theta)$), for the cone angle $\theta \Equ$30$\degree$). The hyperboloid height is taken to be $h \Equ$6.23 pc, dusty disc radius is $D \Equ$3 pc, while angle of dusty disc is $\Delta_{{\rm disc}} \Equ$5$\degree$ (see Fig. 1 in S19). The model is rotated around the y-axis and in the Figs.~\ref{fig:p} and \ref{fig:p1} are presented the projected surfaces to the plane of observer of a given model seen under the different inclination angles (2--90$\degree$). 

 In this analysis we assume that the BLR is flattened and coplanar with the dusty disc, or in other words that the axis of symmetry of a dusty cone is perpendicular to the flattened BLR.

A Monte Carlo simulation with 10000 points in 2D space is performed in order to calculate the surfaces of the projections (in pc$^{2}$) of the model to the plane of observer, for the various angles of inclination (as seen in the Figs.~\ref{fig:p} and \ref{fig:p1}). The resulting surfaces of the hyperboloid and the dusty disc components are given in the Table~\ref{tab2}, in the second and the third column, respectively. In the fourth column, these two surfaces are added. Errors of Monte Carlo surface estimation were estimated to be 0.1729\%.

\begin{table} 
\caption{Modeled surfaces for different inclinations: Inclinations (column 1); the surfaces of the projections of the hyperboloid and dusty disc to the plane of observer under the different inclination angles are in columns 2 and 3 ($S_{{\rm hyp}}$ and $S_{{\rm ddisc}}$); while in column 4 there is $S_{{\rm hyp \Plus ddisc}}$ (added surfaces of hyperboloid and dusty disc). \label{tab2}}
\centering{
\begin{tabular}{|c|c|c|c|}
\hline\hline
 i [$\degree$]&$S_{{\rm hyp}}$[pc$^{2}$]&$S_{{\rm ddisc}}$[pc$^{2}$]& $S_{{\rm hyp \Plus ddisc}}$[pc$^{2}$] \\ 
(1) &(2)&(3)&(4)  \\ \hline 
 2  & 41.73$\pm$0.07     &28.14$\pm$0.05      &  69.87$\pm$0.09   \\ \hline     
 10 & 52.74$\pm$0.09     &27.91$\pm$0.05      &  80.65$\pm$0.10   \\ \hline     
 20 & 64.00$\pm$0.11     &26.98$\pm$0.05      &  90.98$\pm$0.12   \\ \hline     
 30 & 68.61$\pm$0.12     &25.36$\pm$0.04      &  93.97$\pm$0.13   \\ \hline     
 40 & 68.10$\pm$0.12     &23.27$\pm$0.04      &  91.37$\pm$0.13   \\ \hline     
 50 & 66.05$\pm$0.11     &20.74$\pm$0.04      &  86.79$\pm$0.12     \\ \hline     
 60 & 63.23$\pm$0.11     &16.64$\pm$0.03      &  79.87$\pm$0.11    \\ \hline     
 70 & 60.16$\pm$0.10     &12.29$\pm$0.02      &  72.45$\pm$0.10      \\ \hline     
 80 & 55.81$\pm$0.10     &8.19$\pm$0.01       &  64.00$\pm$0.10       \\ \hline       
 90 & 52.33$\pm$0.09     &3.78$\pm$0.006      &  56.11$\pm$0.09    \\ \hline     
 \hline
\end{tabular}}
\end{table} 

Afterwards, we applied some different possible cone models without a disc (since the disc would not change the observed surfaces much for large heights). The expected cone angles for AGNs are $\theta \Equ$30-60$\degree$. Here, adopted cone height is 40 pc (although some observed cone heights are order of magnitude higher). We will show that the shape of the dependence of the projection surface (to the plane of the observer) from $i$ does not depend on the cone hight (while the projection surfaces depend on cone height).

\section{Results} \label{sec:Results}

In Section~\ref{sec:jop} we find the correlations among spectral parameters and $i$ (for the dataset in Tables~\ref{tab1}, \ref{tabsr} and \ref{tab3}). In Sections \ref{sec:Results2} and \ref{sec:depth} we tried to estimate the luminosities calculated using the dusty cone models seen under different $i$, and compare their dependence on $i$ with the dependence of the real luminosities from FWHM(H$\beta$).

\subsection{The correlations of spectroscopic parameters with $i$} \label{sec:jop}

\begin{figure}  
\centering
\rotatebox{0}{
\includegraphics[width=87mm]{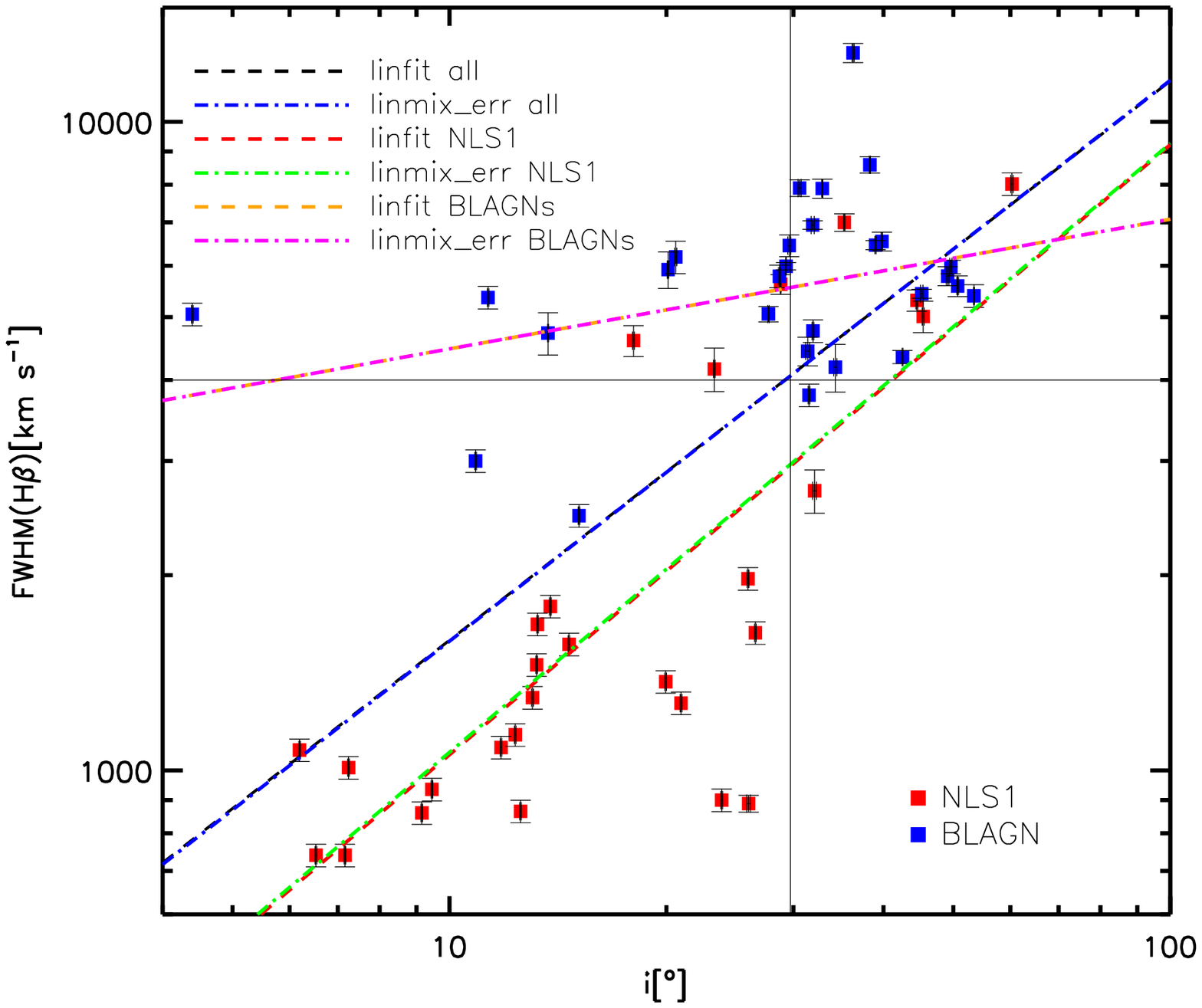}}
\caption{The relations between $i$ and FWHM(H$\beta$) for dataset of Type 1 AGN, for all objects and for NLS1s and BLAGNs separately, including and without errors included (see the legend). The correlations are given in the Table~\ref{tab_fit0}. Horizontal line denotes FWHM(H$\beta$) $\Equ$4000 km s$^{-1}$, and vertical line denotes $i \Equ$29.72$\degree$. Red colour marks NLS1s, while blue colour presents BLAGNs. \label{fig:i_fwhm}}
\end{figure}

The dependence FWHM(H$\beta$)-$i$ is expected to exist, as the projection of the rotational velocity of the matter from accretion disc to the direction to the observer grows with $i$. Therefore, the other implications of this relation are considered. The relation between inclination, $i$ and FWHM(H$\beta$) of Type 1 AGNs dataset \citep[similarly as in][]{Zhang02} is also present here for somewhat different sample (see Fig.~\ref{fig:i_fwhm}). Note that in our previous works we did not consider the errors of parameters (as they were not always available and homogeneous). In order to see how the strength of correlation would be changed, we represent results from correlation analysis both with and without taking into account errors. In Fig.~\ref{fig:i_fwhm} are given linear fitting without errors included ({\sl linfit} in {\sl IDL} for all objects and for NLS1s and BLAGNs separately), together with the fitting using a Bayesian method of linear regression ({\sl linmix\_err} in {\sl IDL}, for all objects and for NLS1s and BLAGNs separately) that accounts for the errors of each parameter \citep[method is described in][]{Kelly07}. All correlations are shown in the Table~\ref{tab_fit0}. One can notice that for BLAGNs there are no trends. In the Bayesian fittings Markov chains were created using the Metropolis-Hastings algorithm. The fitting parameters are similar for these three kinds of fitting (the three pair of lines overlap). 

\begin{table*} 
\caption{Fitting results for Fig.~\ref{fig:i_fwhm}: A -- constant, B -- slope, R -- Pearson correlaton coefficient and P value (without errors included) and: A$_{e}$, B$_{e}$ and R$_{e}$ correlaton coefficient (with errors included), for all objects and for NLS1s and BLAGNs separatelly. As it is shown in the last column, for BLAGNs there are no trends.  \label{tab_fit0}}
\centering{
\begin{tabular}{|c|c|c|c|c|c|c|c|c|}
\hline\hline
        & A             & B            & R       &       P    &       A$_{e}$ &        B$_{e}$ &      R$_{e}$ & Related?\\ \hline
 All    & 0.86$\pm$0.13 &2.34$\pm$0.17 &   0.68  & $<$10$^{-5}$ & 0.86$\pm$0.13 & 2.33$\pm$0.18  &      0.68$\pm$0.07  & yes   \\ \hline     
 NLS1s  & 0.94$\pm$0.15 &2.08$\pm$0.18 &   0.78  & $<$10$^{-5}$ & 0.94$\pm$0.16 & 2.08$\pm$0.20  &      0.81$\pm$0.08  & yes    \\ \hline     
 BLAGNs & 3.45$\pm$0.15 &0.20$\pm$0.10 &   0.36  & 0.062        & 3.45$\pm$0.16 & 0.20$\pm$0.11  &      0.36$\pm$0.18  & no     \\ \hline     
 \hline
\end{tabular}}
\end{table*} 

\begin{figure}  
\rotatebox{0}{
\includegraphics[width=87mm]{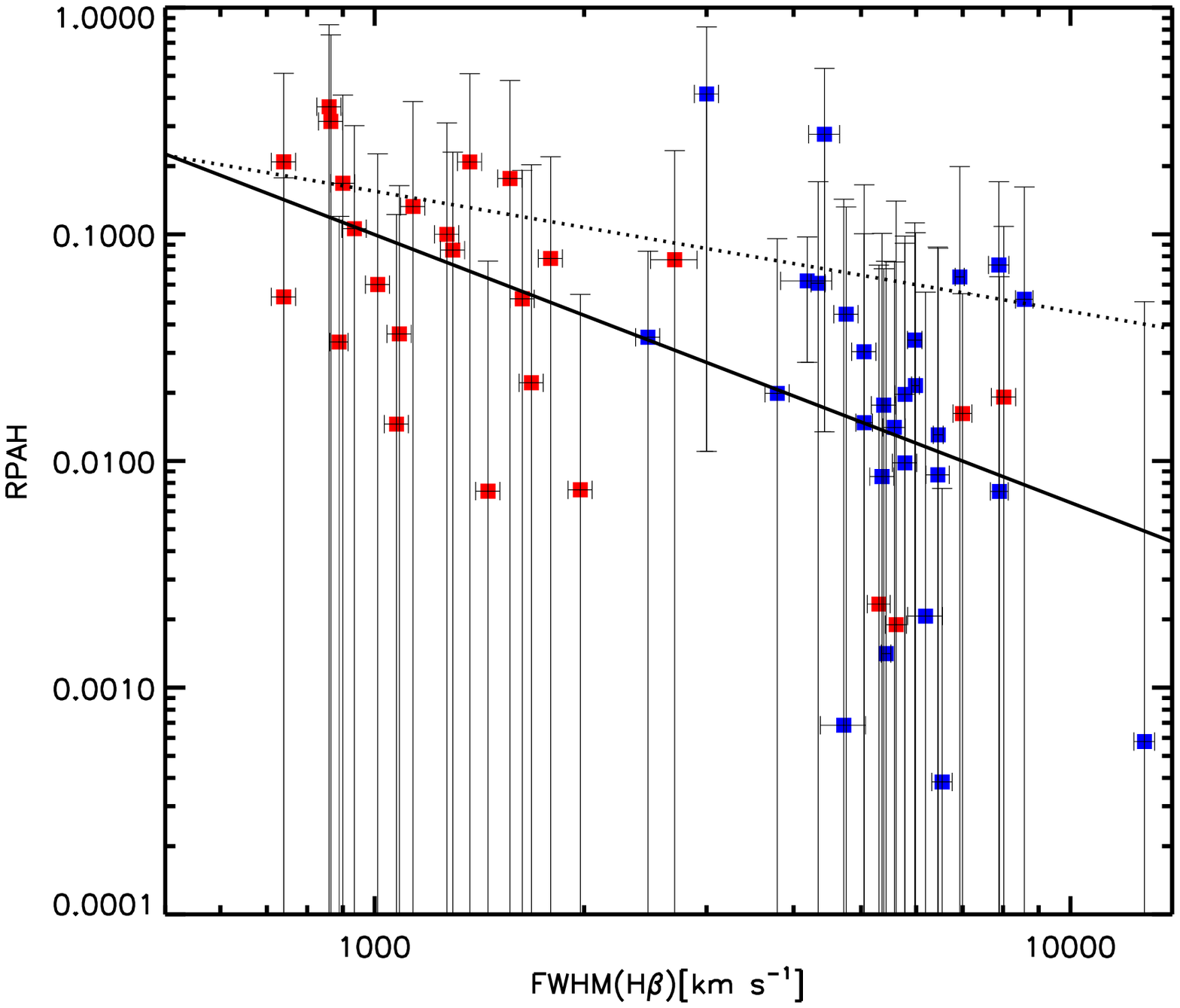}}
\caption{A relation between FWHM(H$\beta$) and RPAH for Type 1 AGNs dataset from this paper. Full line is fit without uncertainties accounted (y$\Equ$(-1.18$\pm$0.24)x$\Plus$(2.54$\pm$0.84); R$\Equ$-0.57; P$<$0.00001), while dashed line is the fit with uncertainties included (y=(-0.53$\pm$0.39)x$\Plus$(0.78$\pm$0.32); R$\Equ$-0.64). The data colour is the same as in Fig.~\ref{fig:i_fwhm}. \label{fig:FWHMHb_RPAH}}
\end{figure}

The known relation of FWHM(H$\beta$) vs. fractional PAH contribution to Spitzer spectra, RPAH \citep{Laki17}, is shown in Fig.~\ref{fig:FWHMHb_RPAH}. The correlation parameters without errors included (full line) and with errors accounted (dotted line) are given on the plot and shown in the Table~\ref{tab_fit2}. One can notice that without including uncertainties, the slope is steeper. La18 found this trend for NLS1 objects, while for BLAGNs it did not exist.

\begin{figure}  
\rotatebox{0}{
\includegraphics[width=75mm]{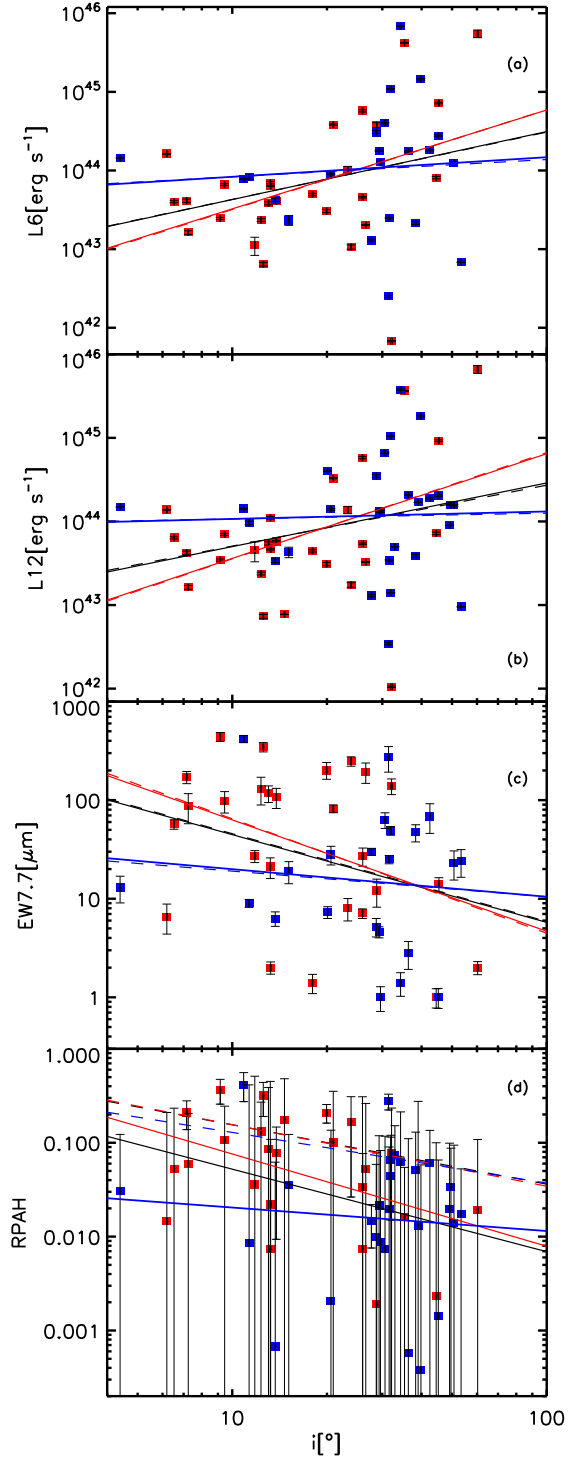}}
\caption{The trends for all objects from the sample: inclination, $i$ vs. L6 (a), L12 (b), EW7.7$\mu$m (c), and RPAH (d). The data colour is the same as in Fig.~\ref{fig:i_fwhm}. Black lines are fittings with all data, red lines are for NLS1s only, while blue lines are for BLAGNs only. Full lines are obtained without using uncertainties, while dashed lines are obtained using uncertainties. \label{fig:i}}
\end{figure}

 In Fig.~\ref{fig:i}, we can notice the weak trends between L6 (a), L12 (b), EW of PAH at 7.7$\mu$m (c) and RPAH (d) with $i$. One can notice that all objects together and NLS1s separately have trends, while BLAGNs separately do not show correlations. All these results are summed in the Table~\ref{tab_fit1}. For Fig.~\ref{fig:i} a, b and c the fitting with and without uncertainties give similar results. La18 showed that all these correlations, FWHM(H$\beta$) vs. L6, L12, the luminosity at 5100 $\angstrom$ (L5100), RPAH, EWPAH exist for NLS1s, while they do not exist for BLAGNs.

\begin{table*} 
\caption{Fitting results for Fig.~\ref{fig:FWHMHb_RPAH}: A -- constant, B -- slope, R -- Pearson correlaton coefficient and P value (without errors included) and: A$_{e}$, B$_{e}$ and R$_{e}$ correlaton coefficient (with errors included), for all objects.   \label{tab_fit2}}
\centering{
\begin{tabular}{|c|c|c|c|c|c|c|c|c|}
\hline\hline
        & A             & B            & R       &       P      &       A$_{e}$ &        B$_{e}$ &      R$_{e}$ & Related?\\ \hline
 All    & 2.54$\pm$0.84 &-1.18$\pm$0.24 &   -0.57& $<$10$^{-5}$ & 0.78$\pm$0.32 &  -0.53$\pm$0.39&      -0.64$\pm$0.68  & yes   \\ \hline     
 \hline
\end{tabular}}
\end{table*} 

\begin{table*} 
\caption{Fitting results for Fig.~\ref{fig:i}: A -- constant, B -- slope, R -- Pearson correlaton coefficient and P value (without errors included) and: A$_{e}$, B$_{e}$ and R$_{e}$ correlaton coefficient (with errors included), for all objects, and for NLS1s and BLAGNs separatelly. As it is shown in the last column, for BLAGNs there are no trends.  \label{tab_fit1}}
\centering{
\begin{tabular}{|c|c|c|c|c|c|c|c|c|c|}
\hline\hline
    &mark  & A             & B             & R       &       P    &       A$_{e}$ &        B$_{e}$ &      R$_{e}$ & Related?\\ \hline 
 All    &a &42.77$\pm$0.54&0.86$\pm$0.40  & 0.299   & 0.036       &42.77$\pm$0.56& 0.86$\pm$0.41  &       0.30$\pm$0.13  & yes    \\ \hline 
 NLS1s  &a &42.25$\pm$0.70&1.26$\pm$0.55  & 0.414   &0.032        &42.24$\pm$0.74& 1.27$\pm$0.57  &       0.46$\pm$0.19  & yes    \\ \hline 
 BLAGNs &a &43.67$\pm$0.96&0.25$\pm$0.67  & 0.084   &0.710        &43.70$\pm$1.00& 0.22$\pm$0.69  &       0.07$\pm$0.22  & no      \\ \hline \hline 
 
 All    &b &42.94$\pm$0.48&0.76$\pm$0.35  &0.285    & 0.033       &42.97$\pm$0.51& 0.73$\pm$0.37  &       0.28$\pm$0.12  & yes   \\ \hline 
 NLS1s  &b &42.30$\pm$0.66&1.25$\pm$0.52  & 0.426   &0.024        &42.28$\pm$0.69& 1.27$\pm$0.54  &       0.46$\pm$0.18  & yes    \\ \hline
 BLAGNs &b &43.94$\pm$0.79&0.09$\pm$0.54  & 0.032   & 0.872       &43.96$\pm$0.81& 0.07$\pm$0.55  &       0.03$\pm$0.20  & no	  \\ \hline  \hline 
 
 All    &c &2.54$\pm$0.51 &-0.89$\pm$0.38 & -0.322   & 0.025      &2.55$\pm$0.53 & -0.89$\pm$0.40 &      -0.34$\pm$0.14  & yes   \\ \hline      
 NLS1s  &c &2.93$\pm$0.69 &-1.13$\pm$0.55 &-0.390     & 0.049      &2.97$\pm$0.71 & -1.16$\pm$0.56 &      -0.43$\pm$0.19  & yes    \\ \hline      
 BLAGNs &c &1.58$\pm$0.86 &-0.28$\pm$0.60  &-0.102   & 0.650      &1.54$\pm$0.89 & -0.26$\pm$0.62 &      -0.10$\pm$0.24  &  no	  \\ \hline \hline 
 
 All    &d &-0.40$\pm$0.47 &-0.88$\pm$0.34 & -0.340   & 0.013      &-0.18$\pm$0.63 & -0.63$\pm$0.47 &      -0.68$\pm$1.20   & yes   \\ \hline       
 NLS1s  &d & -0.14$\pm$0.54 &-0.98$\pm$0.44&-0.420    & 0.034      &-0.15$\pm$1.15 & -0.65$\pm$0.97 &      -0.61$\pm$1.78  & yes    \\ \hline     
 BLAGNs &d & -1.44$\pm$0.89 &-0.25$\pm$0.60  &-0.084 & 0.678      &-0.35$\pm$1.18 & -0.54$\pm$0.82 &      -0.35$\pm$0.78  &  no	  \\ \hline  \hline    

\hline
\end{tabular}}
\end{table*} 

\subsection{The correlations for model of dusty hyperboloid shell compared with spectroscopic correlations} \label{sec:Results2}

\begin{figure}  
\rotatebox{0}{
\includegraphics[width=85mm]{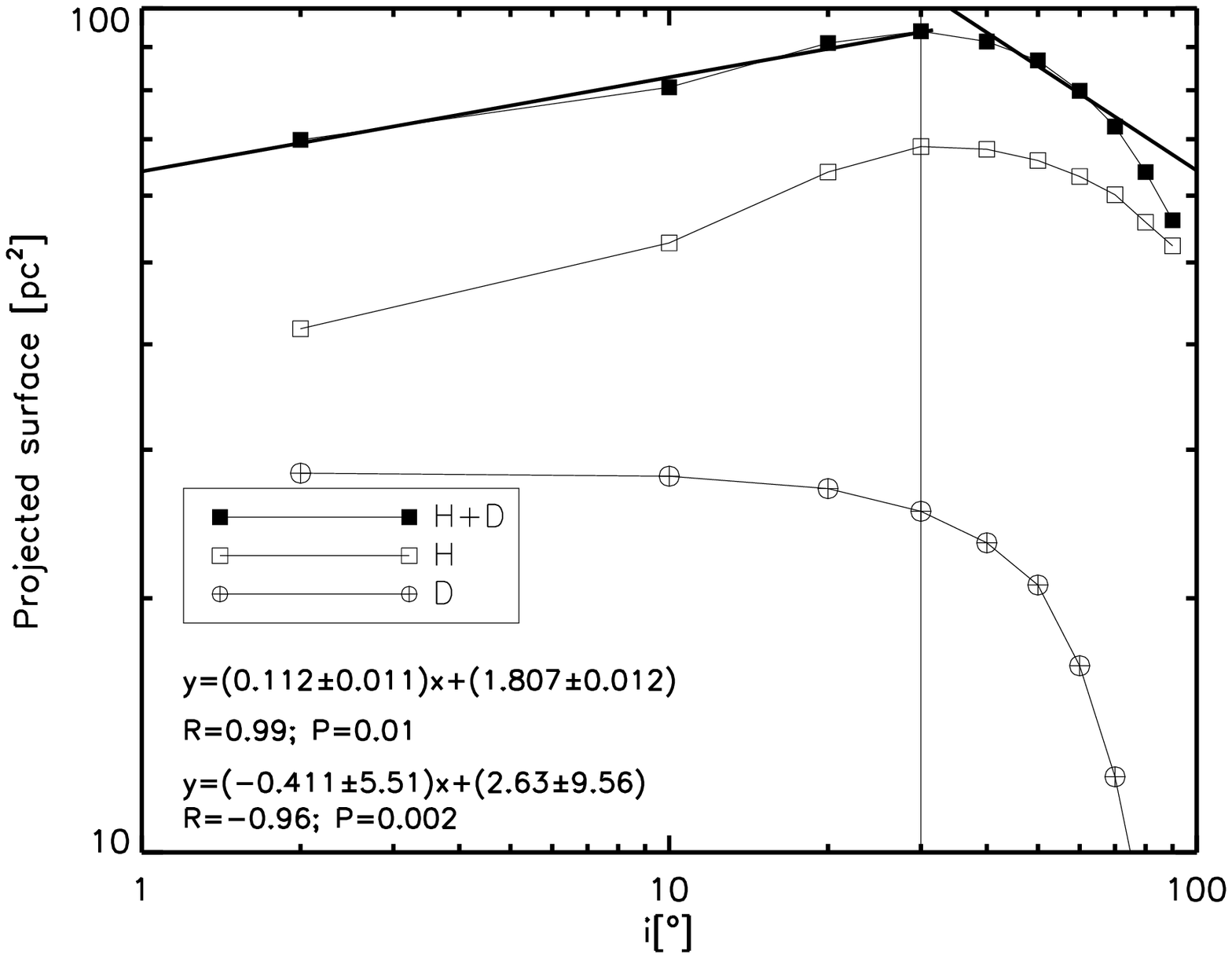}}
\caption{Projected surfaces (to the plane of observer) of hyperboloid (H), disc (D) and added hyperboloid and disc (H$\Plus$D) for the data from the Table~\ref{tab2}. The sum of hyperboloid and dusty disc surfaces increases up to $\sim$30$\degree$ (vertical line), and then it drops. For hyperboloid, projected surface also rises up to $\sim$30$\degree$, and then decline, while for dusty disc it decreases all the time. \label{fig:gla}}
\end{figure}

In Section~\ref{sec:surfaces} we calculated the surfaces of hyperbola and dusty disc projections to the plane of the observer (Table~\ref{tab2}, Figs.~\ref{fig:p} and~\ref{fig:p1}). These projected surfaces in dependence from $i$ are shown in Fig.~\ref{fig:gla}. As one can see from the Table~\ref{tab2} and Fig.~\ref{fig:gla}, for hyperbola part the projection surfaces ($S_{{\rm hyp}}$) grow for inclinations up to $\sim$30$\degree$, and then they decline, while $S_{{\rm ddisc}}$ decrease with the inclination. Their sums, $S_{{\rm hyp \Plus ddisc}}$, also rise for inclinations up to $\sim$30$\degree$, and then they fall. This sum of surfaces (fourth column of Table~\ref{tab2}) should be determining the luminosity of the AGN (as the luminosities are additive), in case that the sheltering is not too high. 

\begin{figure} 
\centering
\rotatebox{0}{
\includegraphics[width=72mm]{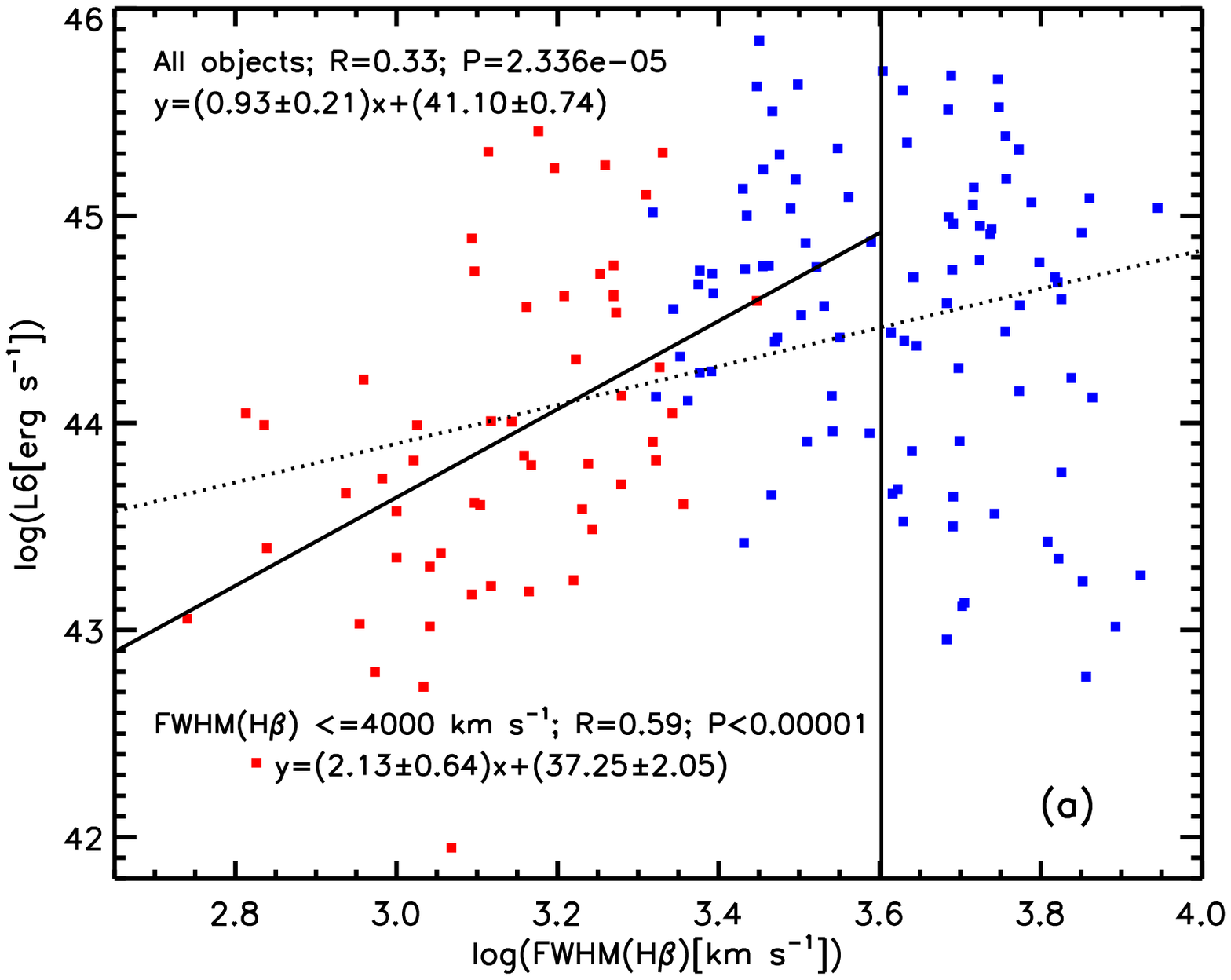}}
\rotatebox{0}{
\includegraphics[width=72mm]{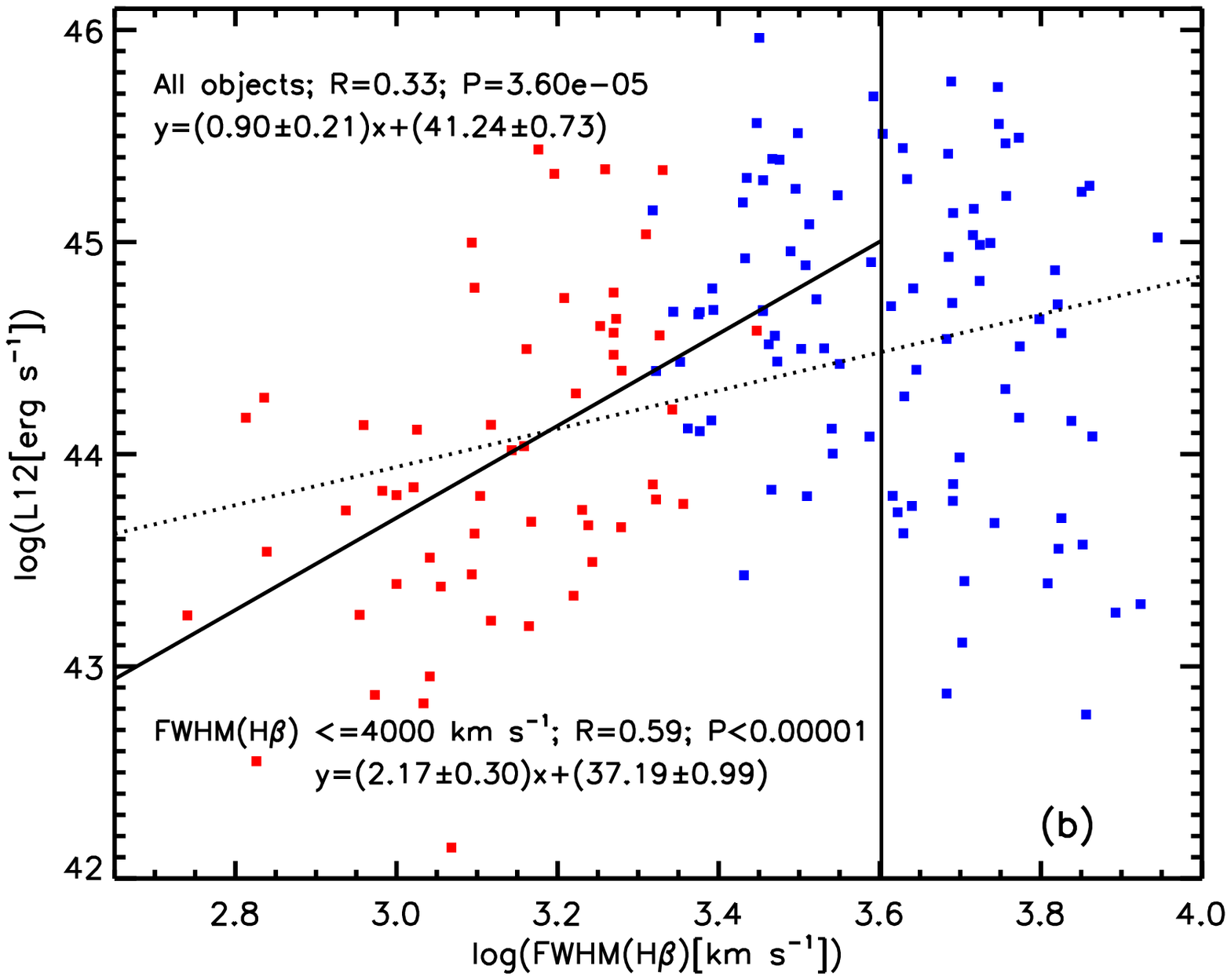}}
\rotatebox{0}{
\includegraphics[width=72mm]{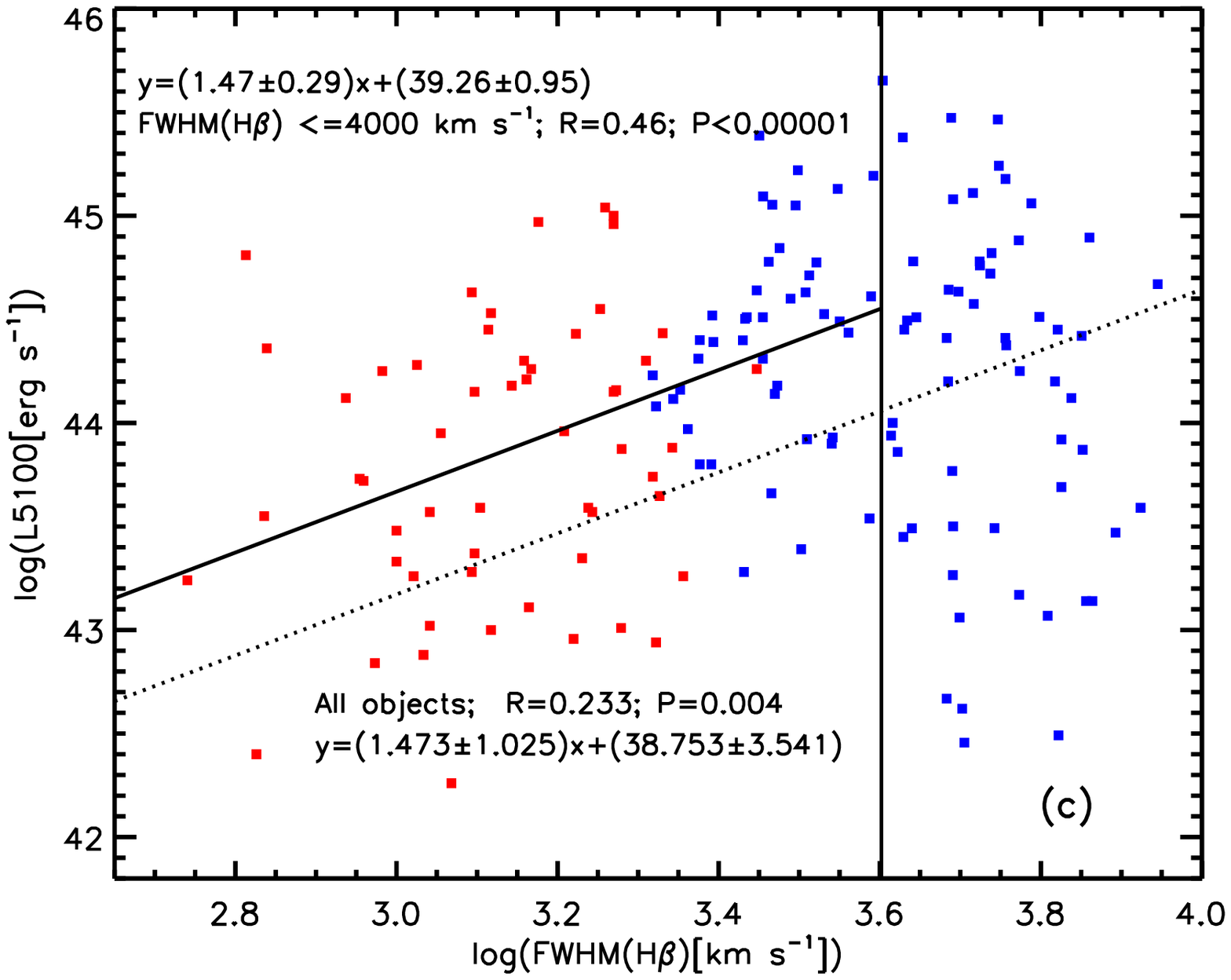}}
\caption{The relations between FWHM(H$\beta$) and L6 (a), FWHM(H$\beta$) and L12 (b) and between FWHM(H$\beta$) and L5100 (c), for the data from La18. The coefficient of correlation (R) and best linear fit are calculated for objects with FWHM(H$\beta$)$<$4000 km s$^{-1}$, left from vertical line (fitted with solid line), since for FWHM(H$\beta$)$>$4000 km s$^{-1}$ there are no trends, and for all objects (fitted with dotted line). The data colour is the same as in Fig.~\ref{fig:i_fwhm}.} \label{fig:La18}
\end{figure}

 $S_{{\rm hyp \Plus ddisc}}$ increases and then decreases with a turning point around 30$\degree$ (Fig.~\ref{fig:gla}). The trends for $i\le$30 $\degree$ and $i\ge$30$\degree$ are given in the Fig.~\ref{fig:gla}. That reminds on the spectroscopic relations shown in Fig.~\ref{fig:La18}, where the three correlations from the data from La18 paper are given, FWHM(H$\beta$) vs. the luminosities at 6 and 12 $\mu$m and the luminosity at 5100 $\angstrom$, (a), (b) and (c), respectively. In La18, it was presented that NLS1s have positive correlations between FWHM(H$\beta$) and aforementioned luminosities, luminosity of broad H$\beta$ line and several MIR coronal lines, while BLAGN do not have, or have weak negative correlations. Also, La18 noticed that RPAH is anticorrelated with FWHM(H$\beta$) for NLS1s (Fig.~\ref{fig:FWHMHb_RPAH}). 
 Note that in Fig.~\ref{fig:La18} (a), (b) and (c) we present the trends for total sample (dotted) line), separately from the trends for objects with FWHM(H$\beta$) $\le$4000 km s$^{-1}$ (full line), which is different than the boundary between NLS1s and BLAGNs, FWHM(H$\beta$)$\Equ$2200 km s$^{-1}$ \citep{Rakshit17a}, for the two reasons. Firstly, it looks like the turning point of these datapoints according to free assessment, secondly because the relation in Fig.~\ref{fig:i_fwhm} suggests that FWHM(H$\beta$)$\Equ$4000 km s$^{-1}$ corresponds to 29.72$\degree \sim$30$\degree$ (which is actually also the turning point for Fig.~\ref{fig:gla}). By chance, the correlations are stronger for this boundary. For the FWHM(H$\beta$)--L6, for FWHM(H$\beta$) $\le$2200 km s$^{-1}$, the trend had Pearson correlation R$\Equ$ 0.51; P$\Equ$ 6.19x10$^{-5}$, while the correlation for FWHM(H$\beta$) $\le$4000 km s$^{-1}$ is shown in Fig.~\ref{fig:La18} (a), and it is somewhat stronger. Similarly, for FWHM(H$\beta$) $\le$2200 km s$^{-1}$, the trend of FWHM(H$\beta$)--L12 had Pearson correlation R$\Equ$0.497; P$\Equ$ 0.00011, while for FWHM(H$\beta$) $\le$4000 km s$^{-1}$the stronger trend is shown in Fig.~\ref{fig:La18} (b). For the FWHM(H$\beta$)--L5100, for FWHM(H$\beta$) $\le$2200 km s$^{-1}$ we did not find a trend, while the trend for FWHM(H$\beta$) $\le$4000 km s$^{-1}$ is shown in Fig.~\ref{fig:La18} (c). The rest of the data (FWHM(H$\beta$) $>$4000 km s$^{-1}$) does not show significant trends.

\begin{figure}  
\rotatebox{0}{
\includegraphics[width=88mm]{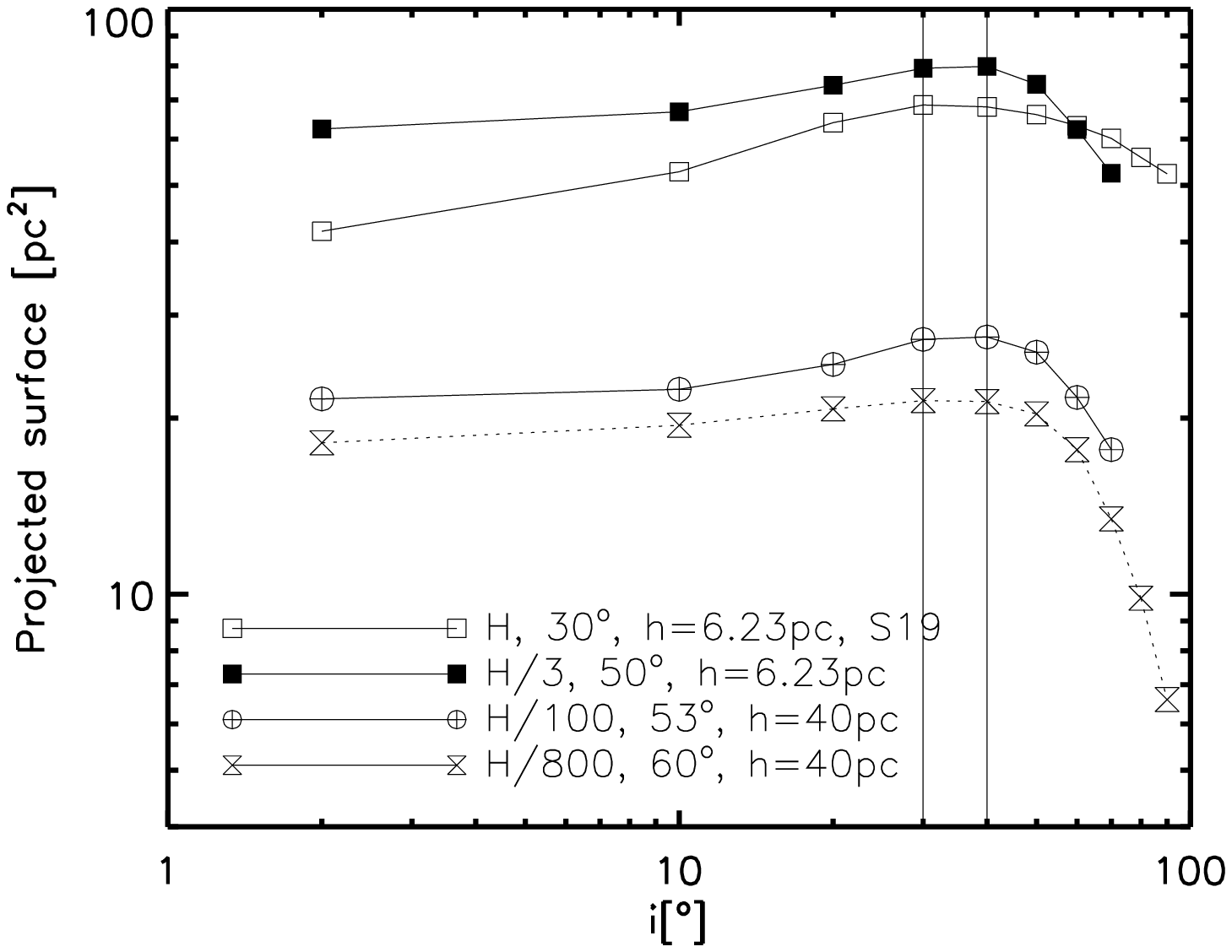}}
\caption{Projected surfaces (to the plane of observer) of hyperboloid (H) for various angles $\theta$ and hyperboloid heights h (see the legend). The real H surfaces are divided to certain numbers (see the legend) in order to be presented in the same plot. Some curves break around 30$\degree$, while others break around 40$\degree$. \label{fig:glaP}}
\end{figure}

In Fig.~\ref{fig:glaP} we show the surfaces of projections (to the plane of the observer) of 4 hyperboloids, two with heights 6.23 pc and two with heights of 40 pc (see the legend), in dependence from $i$. They have $\theta \Equ$ 30, 50, 53 and 60$\degree$. One can notice that the curve shape depends on the angle (steeper rise and shallower drop for smaller angles), but does not depend on the cone height. All curves flatten out around 30-40$\degree$, and afterwards they start decreasing.

\subsection{Including optical depth} \label{sec:depth}

\begin{figure}  
\rotatebox{0}{
\includegraphics[width=88mm]{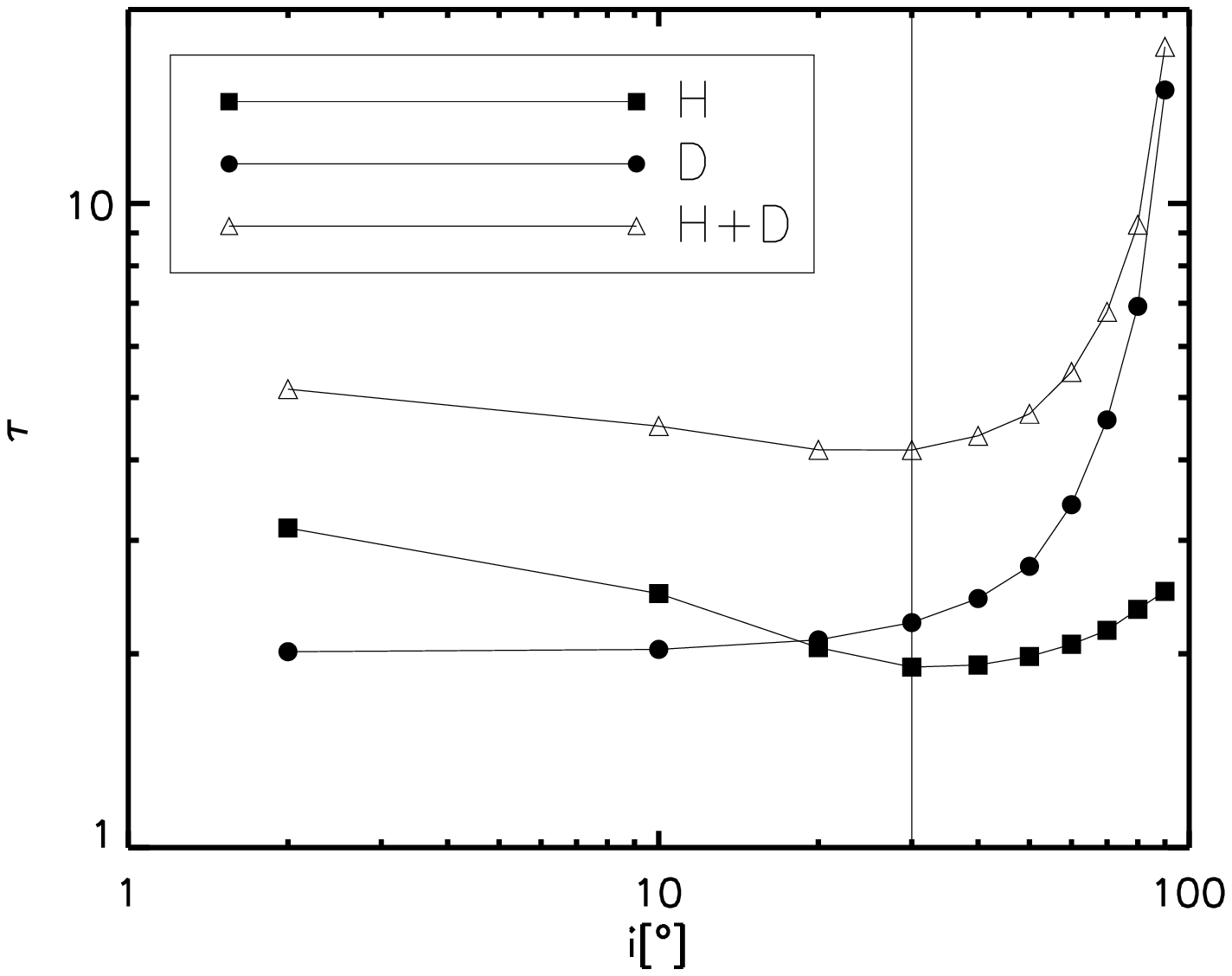}}
\caption{Average estimation of $\tau$ for various inclinations for S19 model, for hyperboloid (H), disc (D) and hyperboloid $\Plus$ disc (H$\Plus$D). \label{fig:tau}}
\end{figure}

\begin{figure}  
\rotatebox{0}{
\includegraphics[width=87mm]{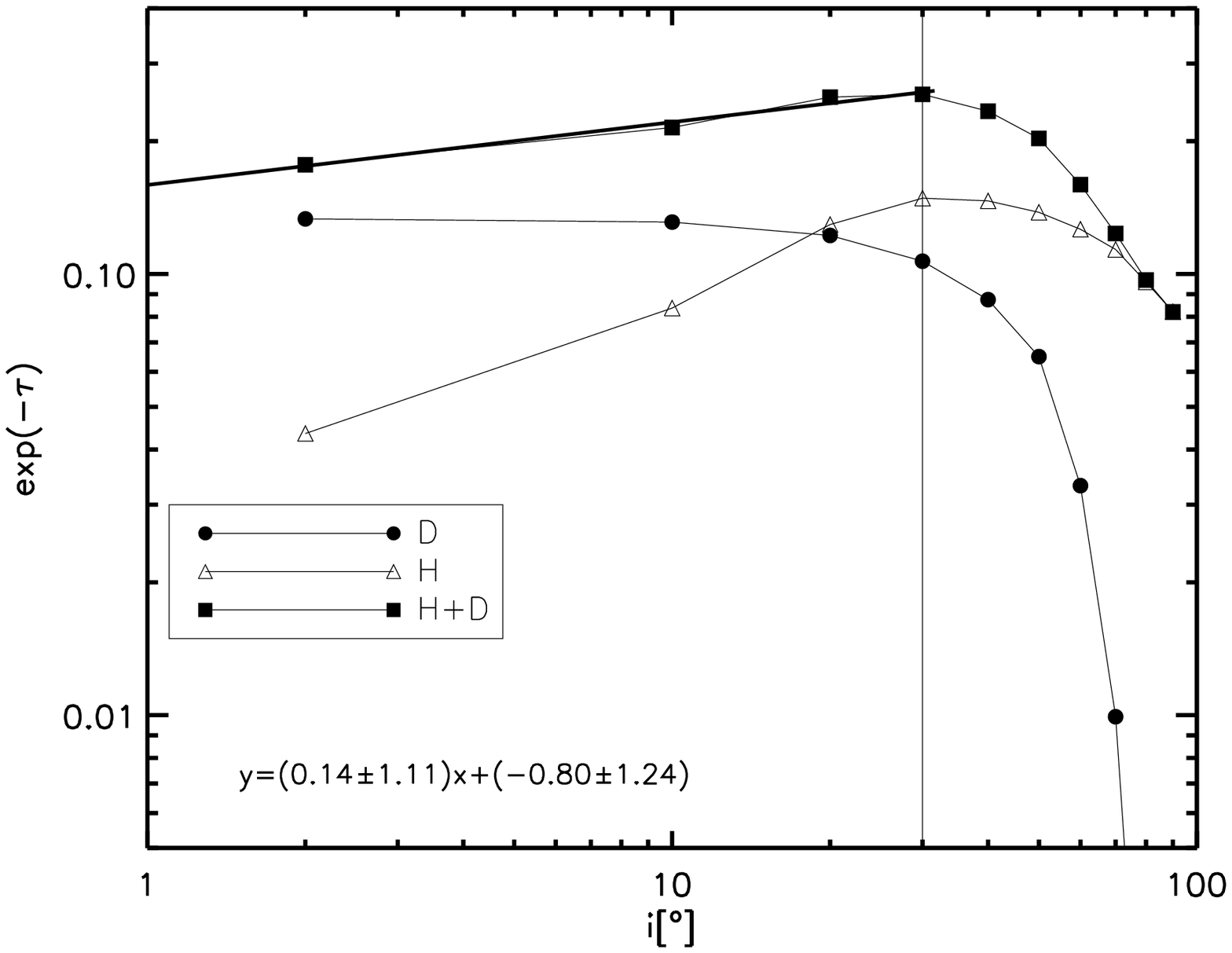}}
\caption{Average estimation of I$\sim$ e$^{-\tau}$ for various inclinations of disc (D), hyperboloid (H), and disc $\Plus$ hyperboloid (H$\Plus$D), for S19 model. The vertical line marks $i \Equ$30$\degree$, where the break for cone occurs. \label{fig:sim}}
\end{figure}

We assume that there is some loss in the MIR radiation due to absorption and scaterring, that for $i\Equ$90$\degree$ the optical depths of hyperboloid and dusty disc are $\tau_{90\degree}^{h} \Equ$2.5 and $\tau_{90\degree}^{d} \Equ$15, respectively (for S19 model), and that the density and the opacity are constant. Here we included the optical depths of hyperboloid and dusty disc in rough approximations. If we define the optical depth, $\tau$ as the height ot the fictive cylinder of the same volume $V$ as   cone/dusty disc, and the base $S$ (projection surface), such that $V$/$S\Equ \tau$, then
\begin{equation}
 \frac{V}{S_{90, {\rm hyp}}} : \tau_{90 \degree}^{h} \Equ \frac{V}{S_{i, {\rm hyp}}} : \tau_{i}^{h} \Rightarrow \frac{\tau_{i}^{h}}{S_{90, {\rm hyp}}} \Equ \frac{\tau_{90}^{h}}{S_{i, {\rm hyp}}},
\end{equation} therefore the optical depths of hyperboloid and dusty disc can be approximated as:

\begin{equation}
 \tau_{i}^{h} \Equ \frac{S_{90 \degree, {\rm hyp}}}{S_{i, {\rm hyp}}} \times \tau_{90 \degree}^{h}
\end{equation} and 
\begin{equation}
 \tau_{i}^{d} \Equ \frac{S_{90 \degree, {\rm ddisc}}}{S_{i, {\rm ddisc}}} \times \tau_{90 \degree}^{d}.
\end{equation} 
The dependence of $\tau$ from $i$ is given in Fig.~\ref{fig:tau}. Including $\tau$, the intensity of the source is approximately estimated with formulas I$_{h} \sim$ e$^{-\tau_{{\rm hyp}}}$ and I$_{d} \sim$ e$^{-\tau_{{\rm ddisc}}}$, see Fig~\ref{fig:sim}. Here, the peak of intensity is again around 30$\degree$. The curve for a cone (Fig~\ref{fig:sim}) is somewhat steeper than the one in Fig.~\ref{fig:gla}, even if $\tau \Equ \tau$/2 would be taken. Also, the curves for cones for other (larger) angles $\theta$ are also getting somewhat steeper when the average optical thickness is included. However, these approximations are rough and the more detailed approach is needed, such as radiative transfer.

\subsection{Comparison with observational datasets} \label{sec:others}

As the correlations that could be found for some sample depend on the characteristics of the objects from that sample (as the range of the cosmological redshift, continuum luminosities, M$_{{\rm BH}}$, broad lines width. etc.), we checked if luminosities-FWHM(H$\beta$) correlations can be found for samples that significantly differ from the dataset from this work. The characteristics of these NLS1 samples and their correlations are given in the Table~\ref{tab_samples}. 

Having in mind other datasets and measurements of FWHM(H$\beta$) vs. various luminosities, we noticed that the luminosities are not necessarilly in the same trend with FWHM(H$\beta$). For example as e.g. Fig. 12 from \citet{Zhou06} (0$<z<$0.8; 40$<$log(LH$\beta$)$<$44), where FWHM(H$\beta$) and FWHM(H$\alpha$) vs. their luminosities are given for NLS1s, objects are confined under the guiding line and their trends are only R$\Equ$0.29 and R$\Equ$0.34, respectively. They explain this by the existence of upper limits in the accretion rate in units of the Eddington rate. On the other hand, \citet{Veron01} obtained stronger correlation for FWHM(H$\beta$) and its luminosity: 0.76, while this relation covers both NLS1 objects with z$<$0.1, 40$<$log(LH$\beta$)$<$44 and BLAGNs with 41$<$log(LH$\beta$)$<$46.

\begin{table*} 
\caption{The characteristics and chosen correlation coefficients for NLS1 data samples from Section~\ref{sec:others}.  \label{tab_samples}}
\centering{
\begin{tabular}{|c|c|c|c|c|c|c|}
\hline\hline
 Sample                 & z & log(Lumin.) [erg s$^{-1}$]  &size & FWHM(H$\beta$)-L(H$\beta$)     &FWHM(H$\beta$)-X lumin. & FWHM(H$\beta$)-L5100 \\ \hline
 this work              & z$<$0.3       & 42$<$L5100<46       &56       &    NA                        &       NA                  & NA                 \\ \hline
  \citet{Zhou06}        & 0$<$z$<$0.8   &40$<$log(LH$\beta$)$<$44 &2000     & 0.29                     &     NA                    & NA                  \\ 
                        &               &41$<$L5100<46            &         &                          &                           &                   \\ \hline
 \citet{Veron01}        & z$<$0.1       &40$<$log(LH$\beta$)$<$46 & 64      & 0.76                     &    NA                     &  NA                \\  \hline                       
 \citet{Kovacevic10}    & z$<$0.7       &43$<$log(L5100)$<$47     & 92      &  0.39                    &     NA                    &  0.33            \\
 FWHM(H$\beta$)$<$2200 km s$^{-1}$ &    &40$<$log(LH$\beta$)$<$44)&         &   P$\Equ$0.00013         &                           &  P$\Equ$0.0015    \\ \hline
 \citet{Rakshit17a}     &z$<$0.8        &42$<$log(L5100)$<$45  &     11101  &        NA                &     NA                    & no trend    \\ \hline
 \citet{Bianchi09}      &z$<$0.4        &41$<$Hard X lum. $<$45   &23     &     NA                   &   Hard: 0.37; Hard and      &        NA        \\ 
                        &               &                           &       &                          &   soft lum. ratio: -0.64   &               \\  \hline
 \citet{Laki18}         &z$<$0.7        &42$<$log(L5100)$<$46    & 64       &  0.734                   &0.471                      & no trend    \\
                        &               &                        &        &P$\Equ$7.9$\times$10$^{-5}$    &P$\Equ$0.004                      &             \\ \hline
 \hline
\end{tabular}}
\end{table*}

\begin{figure}  
\rotatebox{0}{
\includegraphics[width=87mm]{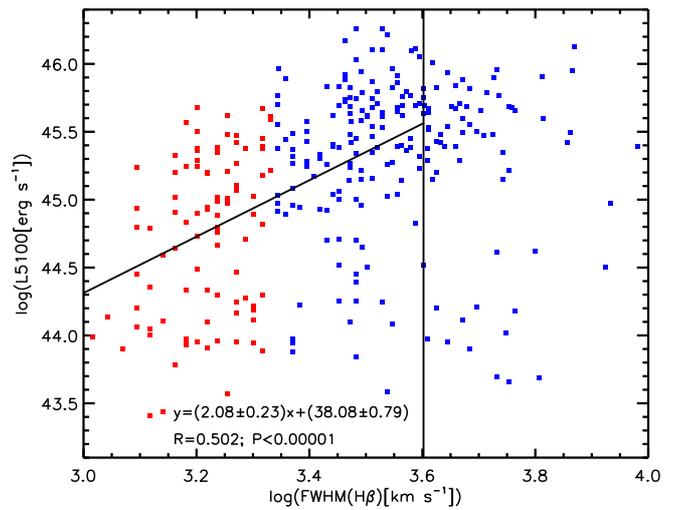}}
\caption{A relation FWHM(H$\beta$)--L5100 made from the data from paper \citet{Kovacevic10}, for the objects where FWHM(H$\beta$)$<$4000 km s$^{-1}$. The data colour is the same as in Fig.~\ref{fig:i_fwhm}.} \label{fig:JK}
\end{figure}

Regarding the dataset from \citet{Kovacevic10} ($z<0.7$; 43$<$log(L5100)$<$47; 40$<$log(LH$\beta$)$<$44), the FWHM(H$\beta$)--L5100 relation from their measurements and trend for FWHM(H$\beta$)$\le$4000 km s$^{-1}$ are given in Fig.~\ref{fig:JK}. Similarly as before, the correlations are higher for FWHM(H$\beta$)$\le$4000 km s$^{-1}$ than for NLS1s only (for FWHM(H$\beta$)$\le$ 2200 km s$^{-1}$, R=0.326, P=0.0015). Analyzing the same sample, \citet{Popovic11} noticed high FWHM(H$\beta$)--L5100 correlation for the objects with high starburst contribution, see their Fig. 7, left.

\begin{figure}  
\rotatebox{0}{
\includegraphics[width=87mm]{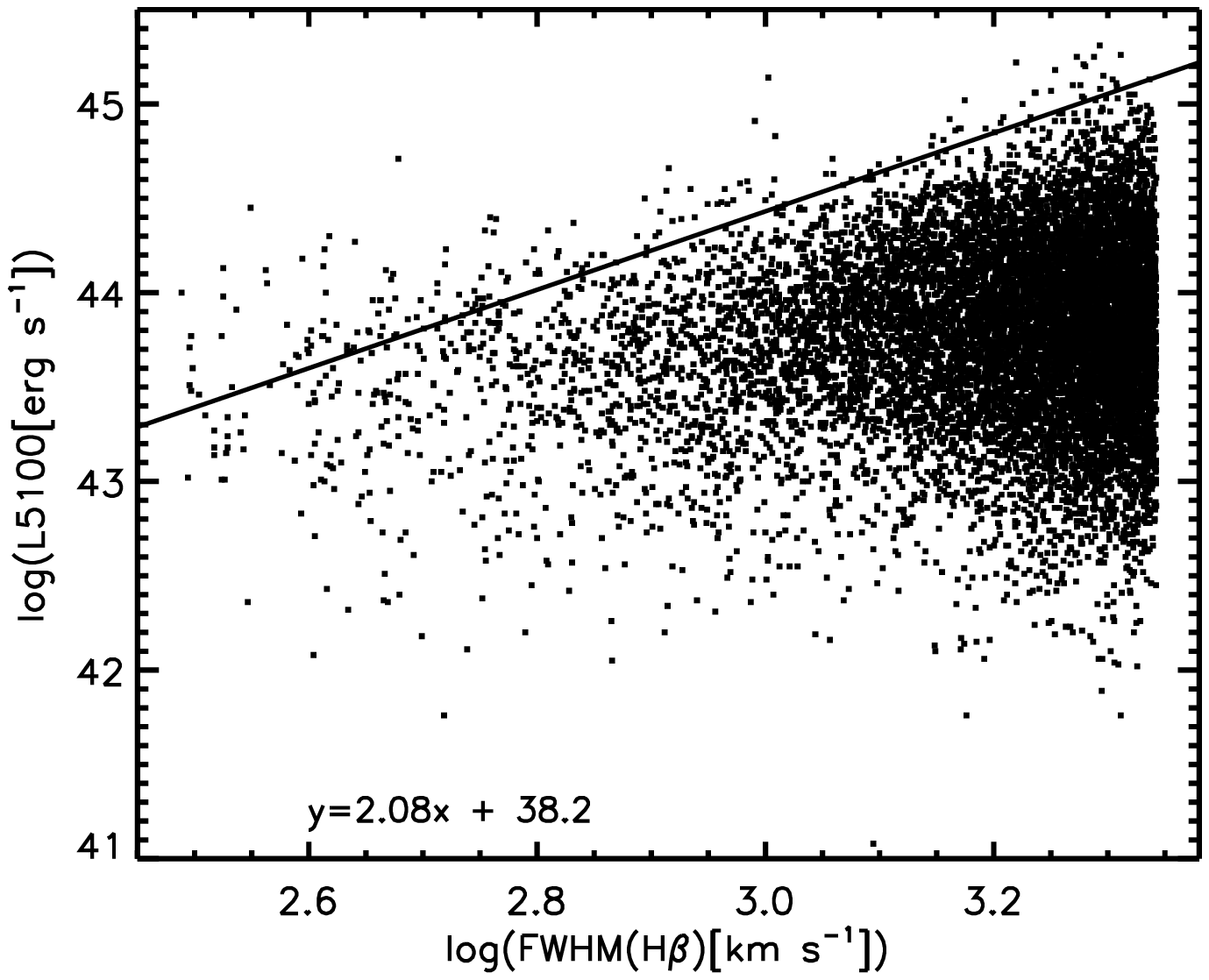}}
\caption{FWHM(H$\beta$)--L5100 plot, made from the catalogue of NLS1s in \citet{Rakshit17a}. \label{fig:rak}}
\end{figure}

The FWHM(H$\beta$)--L5100 relation for NLS1s from \citet{Rakshit17a} ($z<0.8$; 42$<$log(L5100)$<$45) is presented in Fig~\ref{fig:rak}. Here NLS1s do not show the trend with FWHM(H$\beta$) (coefficient R$\Equ$0.142; P$<$10$^{-6}$), but they are approximatelly settled below the line y$\Equ$2.08x $\Plus$ 38.2. In this sample they included only objects with FWHM(H$\beta$)$<$2200 km s$^{-1}$.

A FWHM(H$\beta$) vs. X-ray luminosity plot, described in La18, can be seen in \citet{Bianchi09}, in their Fig.~2. X-ray luminosity--FWHM(H$\beta$) correlation exists for NLS1s only (same like for MIR and optical luminosities in other datasets). The most of objects in the sample have $z\lesssim0.4$.

\section{Discussion} \label{sec:Disc}

\citet{Zhang02} used disc-like BLR model, calculating M$_{\rm BH}$ from the stellar velocity despersion (as this method is not too dependent on the BLR inclination), to obtain the inclination of the BLR, $i$. \citet{Zhang02} noticed that $i$ is dependent on the FWHM(H$\beta$) (see their Fig.~3 and Section~\ref{sec:jop} in this paper). This encourages us to connect aforementioned correlations (FWHM(H$\beta$)--luminosities) with the inclination and perhaps with spectral properties of NLS1s and BLAGNs (discussed in Section~\ref{sec:s1}). In the Section~\ref{sec:s2} we consider the possible connection of FWHM(H$\beta$)--luminosities correlations with the cone model of AGNs.

\subsection{NLS1s, BLAGNs, inclination and spectroscopic correlations}  \label{sec:s1}

The calculation of the inclination angle of BLR using Equation~\ref{eq:eq1} predicts the higher FWHM of broad lines for higher inclinations (Fig.~\ref{fig:i_fwhm}), as the projections of the velocities of the matter around BHs to the direction to the observer are higher for objects with higher $i$. Therefore, NLS1s could be the same objects as BLAGNs, but observed under the lower inclination \citep{Nagao00,Zhang02}. \citet{Zhang02} found that $i$ are significantly higher for BLAGNs than for NLS1s and that $i$ for NLS1s rarely exceeds 30$\degree$.

Various datasets such as La18, \citet{Kovacevic10,Veron01,Zhou06,Bianchi09,Jarvela15,Laki18} show FWHM(H$\beta$)-luminosities correlations for NLS1s (Section~\ref{sec:others}, Table~\ref{tab_samples}). These various datasets are similar in the redshift and luminosity range, which means that these correlations exist for that type of objects, and that they are not accidentally biased by some sample choosing. 

Lower luminosities for NLS1s than for BLAGNs are shown in La18 and references therein. In Fig.~\ref{fig:La18}, it seems that FWHM-luminosities trends do not exist only for NLS1s (FWHM(H$\beta$) $\lesssim$2200 km s$^{-1}$ and $i \lesssim$15$\degree$ -- according to relation in Fig.~\ref{fig:i_fwhm}), like it was thought in La18, but up to FWHM(H$\beta$) $\lesssim$4000 km s$^{-1}$, where $i \lesssim$30-40$\degree$; as relations for FWHM(H$\beta$) $\lesssim$4000 km s$^{-1}$ are stronger than only for NLS1s. Similar is with FWHM(H$\beta$)-L5100 relation for \citet{Popovic11} dataset. That could mean that these correlations may not exist due to NLS1-BLAGN differences, but as a consequence of some other reason, such as possibly the geometry and $i$. As it can be seen in spectroscopic measurements in Fig.~\ref{fig:i} a and b, the MIR luminosities have weak trend (R$\Equ$0.28-0.30) with calculated $i$ for all data. Since FWHM(H$\beta$) increases with the inclination (Fig.~\ref{fig:i_fwhm}), and since FWHM(H$\beta$) is correlated with luminosities only for objects with FWHM(H$\beta$)$\lesssim$4000 km s$^{-1}$ (Fig.~\ref{fig:La18}), which corresponds to the $i \sim$30$\degree$ (Fig.~\ref{fig:i_fwhm}), the trend $i$--luminosities is expected to exist for objects with $i\lesssim30-40 \degree$ in particular. However, this is not the case; the trends in Fig.~\ref{fig:i} become insignificant if the objects with $i >$ 30-40$\degree$ are excluded from the fit. If the objects with $i >$40$\degree$ are excluded, the correlation parameters for L6 are: P$\Equ$0.23; R$\Equ$0.61 and for L12 they are: R$\Equ$0.23; P$\Equ$0.12. The correlation may be lost because of small dataset, as the poor correlation could be lost if number of data is lower.  

The separation of objects to the ones with FWHM(H$\beta$)$\lesssim$4000 km s$^{-1}$ and others that have FWHM(H$\beta$)$\gtrsim$4000 km s$^{-1}$ might have the similarity with the separation to Populations A and B from \citet{Sulentic00,Sulentic09}. According to \citet{Sulentic09} 50-60\% of of radio quiet sources have FWHM(H$\beta$)$<$4000 km s$^{-1}$, show blueshift and asymmetry of high ionisation lines, strong FeII emission and a soft X-ray excess (so called Population A); while almost all radio loud sources have FWHM(H$\beta$)$>$4000 km s$^{-1}$, weak FeII emission, and lack of a soft X-ray excess (Population B).

Although NLS1s and BLAGNs could have the same structure \citep{Zhang02,Decarli08}, some authors claim that NLS1s may be in early stage of evolution compared to BLAGNs \citep{Mathur2001}, with lighter M$_{\rm BH}$s \citep{Peterson00,Komossa07}. In Fig.~\ref{fig:FWHMHb_RPAH}, similarly as in Fig.~\ref{fig:i} d and e ($i$-EW7.7 and $i$-RPAH), the known FWHM(H$\beta$)-RPAH anticorrelation is shown (present only for the NLS1s; La18). PAHs are more pronounced for the low FWHMs and inclination angles. RPAH is also known to be anticorrelated with M$_{\rm BH}$ and possibly destroyed close to AGN \citep[e.g.][and references therein]{Laki17}, thus the reason for higher RPAH in NLS1s (small $i$) can be lighter M$_{\rm BH}$ in NLS1s. NLS1s have more dusty spiral arms and bars \citep{Deo06}, where PAHs may be located \citep{Siebenmorgen04}, since PAHs are star formation tracer \citep{Peeters04,Shipley16}. Star formation occurs in spiral arms \citep{Baade63}, and spiral arms and bars may be more visible because of low $i$. Afterwards, star formation may drive mass inflow into AGNs, supporting BH growth \citep{Alexander12} and NLS1s have higher accretion rate. There may be some connection between the dusty cone geometry and the starbursts, so that the starburst are more apparent for low inclinations.

\begin{figure}  
\rotatebox{0}{
\includegraphics[width=87mm]{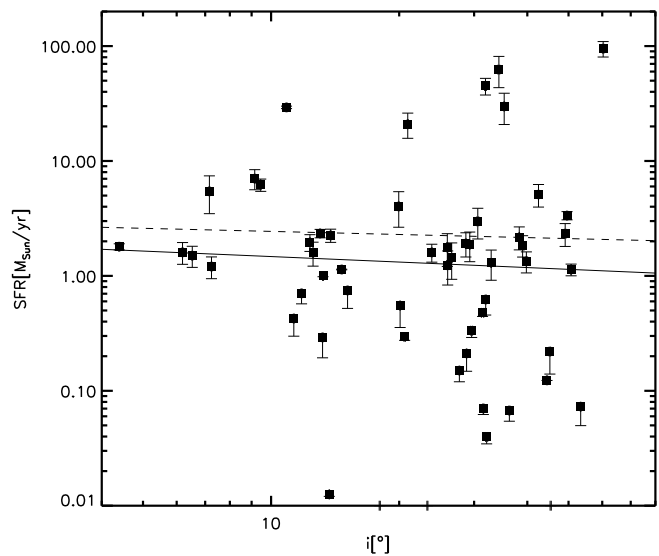}}
\caption{Inclination--SFR plot for sample from this work. The full line is fit without uncertainties, while the dashed line is fit with included errors. \label{fig:sfr}}
\end{figure}

In order to check if star formation rate (SFR) is stronger for lower $i$ (since there RPAH is higher), we used the equation from \citet{Sargsyan09}:
\begin{equation}  \label{eq:sfr}
 log \left[ SFR \left(PAH \right) \right] \Equ log \left[ \nu L_{\nu} \left( 7.7 \mu m \right) \right] \Minus 42.57 \pm 0.2,
\end{equation} to find SFR. There, flux density at the peak of the 7.7 $\mu$m feature is used to find luminosity at 7.7 $\mu$m, $\nu L_{\nu}$(7.7$\mu$m). In Fig.~\ref{fig:sfr} we showed $i$ vs $SFR \left(PAH \right)$ estimated from the Equation~\ref{eq:sfr}. The fitting results without errors included and the ones accounted for the errors are given in the Table~\ref{t:sfr} and the best fits are shown in the Fig.~\ref{fig:sfr}. As one can see from Fig.~\ref{fig:sfr}, there is no obvious correlation between $i$ and SFR, although the sample is small. On the other hand, SFR depends on PAH luminosity, which is different parameter than the PAH contribution in the spectra (RPAH), that is found to be stronger at low inclinations (Fig.~\ref{fig:i} c, d).

\begin{table*} 
\caption{Fitting results for Fig.~\ref{fig:sfr}: A -- constant, B -- slope, R -- Pearson correlaton coefficient and P value (without errors included) and: A$_{e}$, B$_{e}$ and R$_{e}$ correlaton coefficient (with errors included), for all objects. There is no correlation.   \label{t:sfr}}
\centering{
\begin{tabular}{|c|c|c|c|c|c|c|c|c|}
\hline\hline
        & A             & B            & R       &       P      &       A$_{e}$  &        B$_{e}$ &    R$_{e}$ & Related?\\ \hline
 All    & 0.33$\pm$0.54 &-0.16$\pm$0.40 &  -0.06 & 0.69         &  0.49$\pm$0.80 &  -0.09$\pm$0.57&   -0.03$\pm$0.21  & no   \\ \hline     
 \hline
\end{tabular}}
\end{table*} 

\subsection{FWHM(H$\beta$)-luminosity correlations and cone model of AGN}   \label{sec:s2}

Firstly, we assumed a model consisting of the hyperboloid shell and a thin dusty disc, from S19. Seeing an AGN in different angles, the AGN luminosity should be dependent on the surface of the object that the observer is seeing (due to dust obscuration). The two observed surfaces are added and the total surface, $S_{{\rm hyp\Plus ddisc}}$ is obtained. Both $S_{{\rm hyp\Plus ddisc}}$ and $S_{{\rm hyp}}$ increase up to 30$\degree$, and then they drop (Fig.~\ref{fig:gla}), which resembles to the turning point in relations in Fig.~\ref{fig:La18}, but with $\sim$10 times lower slope. However, when the optical depth, $\tau$, in shape $I\sim e^{-\tau}$ is included, the slope for the cone becomes somewhat steeper (Fig.~\ref{fig:sim}). This break that happens both in Fig.~\ref{fig:gla} and for real data (Fig.~\ref{fig:La18}) offers one possible explanation for FWHM(H$\beta$)--luminosities relations. 

Similar results are obtained when the cones of different dimensions and $\theta$ are considered. For all angles $\theta$, thicknesses and heights, the break is always at 30-40$\degree$, while the slopes depend on $\theta$ (Fig.~\ref{fig:glaP}). The smaller angle $\theta$ is, the higher dependence of projection surface to the plane of the observer from inclination (steeper slope) is for $i \lesssim$30$\degree$ and the shallower slope is for $i \gtrsim$30$\degree$.  

Another possible explanation for FWHM(H$\beta$)--luminosities relations is that perhaps low-mass AGNs (AGNs with low luminosities) are not seen in large angles, but only in small angles, since these systems are not confined enough, they are younger objects, their dust may be more likely located in cones and they may be hard to notice at higher $i$ because of the obscuration \citep{Popovic18}. That is one possible selection effect, but we are not aware if there are some other selections that could possibly hide some objects. Together or separately, dependence of luminosities on the observed object surface and/or a lack of light M$_{{\rm BH}}$s for large $i$ may be cause of mentioned correlations.   

It could be expected that the decrease of the luminosity for higher angles ($\gtrsim$30$\degree$) may be also due the observing through the dusty cone or torus or catching more galactic noise. Also, the disagreement between Figs.~\ref{fig:gla} and \ref{fig:sim} with Fig.~\ref{fig:La18} for $i \gtrsim$30$\degree$, FWHM(H$\beta$)$\gtrsim$4000 km s$^{-1}$ (in Fig.~\ref{fig:La18} the trends are missing) may be due to galactic noise and irregularities. 

Luminosities of lines and continuum at different wavelengths are usually strongly correlated in the datasets of various AGNs, and that could offer the explanation for mentioned luminosity trends. The optical luminosities (L5100) should behave similar as MIR luminosities (L6 and L12). Afterall, MIR cones are spread to the much larger distances ($\sim$ 100 pc), while optical data (L5100 and FWHM(H$\beta$)) come from the BLR which has a size of $\sim$10$^{-2}$ pc \citep{Hawkins07}. The size of MIR cones could be a measurement of its AGN power and probably M$_{\rm BH}$.

Finally, L5100--FWHM(H$\beta$) correlation could be connected with starbursts, as it was suggested by \citet{Popovic11}, who found strong relation (R=0.8), for starburst dominated objects (see their Fig. 7). However, starburst dominated objects are often NLS1s, and therefore geometry could still be the explanation. It is important to do further studies. 

\section{Conclusions} \label{sec:Conclusion}

We explored the influence of calculated inclination ($i$) to the luminosities of the Type 1 AGNs using 1) a sample model of dusty cone shells, and/or 2) the selection effect by the BH mass. Inclinations of AGNs are found from optical data. Since $i$ and FWHM(H$\beta$) are correlated and since FWHM(H$\beta$) have the trends with optical, MIR and X-ray luminosities, $i$ should also have trends with these luminosities for $i\lesssim$30-40$\degree$. Here, it is noticed that FWHM(H$\beta$)-luminosities correlations cover broader range, FWHM(H$\beta$)$\lesssim$4000 km s$^{-1}$ than it was previously thought (FWHM(H$\beta$)$<$2200 km s$^{-1}$; La18). 

The inclination obtained from spectroscopic data vs. luminosity trends for AGNs are similar with the inclination--S trend (S is surface of the observed cone model), because S should determine luminosity. Both of these relations break around 30$\degree$. When the optical depths are included, the slope inclination--S is deeper. That could mean that the dusty cone geometry is the cause of FWHM(H$\beta$)-luminosities correlations. 

The alternative explanation for inclination--luminosity relations may be the selection effect by M$_{\rm BH}$ (low-luminosity AGNs with lighter M$_{\rm BH}$ may be less confined, material spread in cones, and these objects may not be seen at large FWHMs).   

From our investigations we can point out following:
\begin{itemize}
 \item Possibly, $i$-luminosities and $i$--RPAH (FWHM(H$\beta$)-luminosities and FWHM(H$\beta$)--RPAH) relations for FWHM(H$\beta$)$\lesssim$4000 km s$^{-1}$ are consequence of the dusty cone geometry of an AGN and the angle of view.
 \item Beside differencies in BH mass, (NLS1s have lighter black hole mass than BLAGNs), inclination of the dusty conical AGN geometry may also be the cause of differences between NLS1s and BLAGNs, where NLS1s may be seen in lower inclination angles. 
 \item Finding the inclinations and the making AGN models may help in understanding Type 1 and 2 AGNs and possibility of unification scheme. 
\end{itemize}

\section*{Acknowledgments}
This work is supported by the Ministry of Education, Science and Technological Development of Serbia, the number of contract is 451-03-68{/}2020-14{/}200002. The Cornell Atlas of Spitzer/IRS Sources (CASSIS) is a product of the Infrared Science Center at Cornell University, supported by NASA and JPL. Much of the analysis presented in this work  was done with TOPCAT (\url{http://www.star.bris.ac.uk/\~mbt/topcat/}), developed by M. Taylor. 

\section*{Data Availability}

The data underlying this article are available in the article and in its online supplementary material. The data underlying this article are compiled from the literature, calculated from the literature data or measured from the spectra that are available to the public (CASSIS database).





\bsp	
\label{lastpage}
\end{document}